\RequirePackage{ifpdf,xcolor,tikz}
\documentclass[hyper,letterpaper]{JHEP3}
\pdfoutput=1
\usepackage{amsmath,amssymb,amsfonts,bm,empheq}
\usepackage{cite}
\usepackage{nicefrac}
\usepackage[enableskew,vcentermath]{youngtab}
\usepackage{epstopdf}
\usepackage{url}
\usepackage{float}
\newtheorem{proposition}{Proposition}[section]
\newtheorem{lemma}[proposition]{Lemma}

\newtheorem{conjecture}[proposition]{Conjecture}

\def\printname#1{
        \if\draft y
                \smash{\makebox[0pt]{\hspace{-0.5in}
                        \raisebox{8pt}{\tt\tiny #1}}}
        \fi
}

\newlength{\standardunitlength}
\catcode`\@=11
\long\def\@makecaption#1#2{%
     \vskip 10pt

\setbox\@tempboxa\hbox{
       \small\sf{\bfcaptionfont #1. }\ignorespaces #2}%
     \ifdim \wd\@tempboxa >\captionwidth {%
         \rightskip=\@captionmargin\leftskip=\@captionmargin
         \unhbox\@tempboxa\par}%
       \else
         \hbox to\hsize{\hfil\box\@tempboxa\hfil}%
     \fi}
\font\bfcaptionfont=cmssbx10 scaled \magstephalf
\newdimen\@captionmargin\@captionmargin=2\parindent
\newdimen\captionwidth\captionwidth=\hsize
\catcode`\@=12

\newdimen\tableauside\tableauside=1.0ex
\newdimen\tableaurule\tableaurule=0.4pt
\newdimen\tableaustep
\def\phantomhrule#1{\hbox{\vbox to0pt{\hrule height\tableaurule width#1\vss}}}
\def\phantomvrule#1{\vbox{\hbox to0pt{\vrule width\tableaurule height#1\hss}}}
\def\sqr{\vbox{%
\phantomhrule\tableaustep
\hbox{\phantomvrule\tableaustep\kern\tableaustep\phantomvrule\tableaustep}%
\hbox{\vbox{\phantomhrule\tableauside}\kern-\tableaurule}}}
\def\squares#1{\hbox{\count0=#1\noindent\loop\sqr
\advance\count0 by-1 \ifnum\count0>0\repeat}}
\def\tableau#1{\vcenter{\offinterlineskip
\tableaustep=\tableauside\advance\tableaustep by-\tableaurule
\kern\normallineskip\hbox
    {\kern\normallineskip\vbox
      {\gettableau#1 0 }%
     \kern\normallineskip\kern\tableaurule}%
  \kern\normallineskip\kern\tableaurule}}
\def\gettableau#1 {\ifnum#1=0\let\next=\null\else
  \squares{#1}\let\next=\gettableau\fi\next}

\newcommand{\bea}{\begin{eqnarray}}
\newcommand{\eea}{\end{eqnarray}}
\newcommand{\be}{\begin{equation}}
\newcommand{\ee}{\end{equation}}

\def\kraj{\hfill\rule{6pt}{6pt}}

\newcommand\catalannumber[3]{
  \fill[white]  (#1) rectangle +(#2,#2);
  \fill[fill=gray!25]
  (#1)
  \foreach \dir in {#3}{
    \ifnum\dir=0
    -- ++(0,1)
    \else
    -- ++(1,0)
    \fi
  } |- (#1);
  \draw[help lines] (#1) grid +(#2,#2);
  \draw[dashed] (#1) -- +(#2,#2);
  \coordinate (prev) at (#1);
  \foreach \dir in {#3}{
    \ifnum\dir=0
    \coordinate (dep) at (0,1);
    \else
    \coordinate (dep) at (1,0);
    \fi
    \draw[line width=2pt,-stealth] (prev) -- ++(dep) coordinate (prev);
  };
}

\title{Knots-quivers correspondence}

\author{Piotr Kucharski$^{1}$, Markus Reineke$^{2}$, Marko Sto$\check{\text{s}}$i$\acute{\text{c}}$$^{3,4}$, and Piotr Su{\l}kowski$^{1,5}$
\\
$^1$ Faculty of Physics, University of Warsaw, ul. Pasteura 5, 02-093 Warsaw, Poland \\
$^2$ Faculty of Mathematics, Ruhr-Universit\"{a}t Bochum, Universit\"{a}tsstrasse 150, 44780 Bochum, Germany \\
$^3$ CAMGSD, Departamento de Matem\'atica, Instituto Superior T\'ecnico,
Av. Rovisco Pais, 1049-001 Lisboa, Portugal  \\
$^4$ Mathematical Institute SANU, Knez Mihailova 36, 11000 Beograd, Serbia \\
$^5$ Walter Burke Institute for Theoretical Physics, California Institute of Technology, Pasadena, CA 91125, USA 
}

\abstract{We introduce and explore the relation between knot invariants and quiver representation theory, which follows from the identification of quiver quantum mechanics in D-brane systems representing knots. We identify various structural properties of quivers associated to knots, and identify such quivers explicitly in many examples, including some infinite families of knots, all knots up to 6 crossings, and some knots with thick homology. Moreover, based on these properties, we derive previously unknown expressions for colored HOMFLY-PT polynomials and superpolynomials for various knots. For all knots, for which we identify the corresponding quivers, the LMOV conjecture for all symmetric representations (i.e. integrality of relevant BPS numbers) is automatically proved.
\\
\\
\\
\\
\\
\\
\\
\\
\\
\\
\\
{\tt CALT-2017-040}}

\begin{document}

\tableofcontents


\newpage

\section{Introduction}    \label{sec-intro}

BPS states in supersymmetric field theories and string theory have remarkable properties, which have been actively studied in last decades. In this paper we consider BPS states that arise in D-brane systems, which encode properties of knots. The counting of these states leads to invariants of knots, referred to as Labastida-Mari{\~n}o-Ooguri-Vafa (LMOV) invariants (or Ooguri-Vafa invariants). On the other hand, dimensional reduction of such brane systems is expected to lead to a description in terms of a supersymmetric quiver quantum mechanics. In this paper we argue that this is indeed the case, and in consequence properties of BPS states are encoded in the data of moduli spaces of quiver representations, which leads to intriguing relations between knots and quivers. We presented a general idea of this correspondence in \cite{Kucharski:2017poe}. Now, in this paper, we explain more details of the identification between knot invariants and quiver moduli spaces, which enables us to identify explicitly relevant quivers for many knots, including some infinite families of knots, all knots with up to 6 crossings, some knots with thick HOMFLY-PT homology, etc. Understanding structural properties of generating series of knot polynomials also enables us to derive previously unknown expressions for colored HOMFLY-PT polynomials and superpolynomials (and their quadruply-graded generalizations) for several knots. More importantly, our correspondence relates generating series of colored HOMFLY-PT polynomials to motivic Donaldson-Thomas (DT) invariants, which then leads to the proof of the famous integrality of LMOV invariants, conjectured in \cite{OoguriV,Labastida:2000zp,Labastida:2000yw}. We also discuss many other consequences of the relation between BPS states, knots and quivers.

Both types of invariants mentioned above, i.e. LMOV invariants of knots and motivic Donaldson-Thomas invariants of quivers, are defined through factorization of some generating series. Our results, in particular the proof of the LMOV conjecture, follow from the identification of these series, which physically amounts to the identification of the corresponding BPS states. In case of knots, the series in question is the generating series of colored HOMFLY-PT polynomials, and it arises as the expectation value of the Ooguri-Vafa operator. This operator characterizes the system of branes, which provide topological string theory realization of observables in Chern-Simons theory. This system consists of $N$ A-model lagrangian branes wrapping $S^3$ in the deformed conifold $T^*S^3$ Calabi-Yau geometry, and intersecting -- along a knotted curve -- an additional set of lagrangian branes \cite{OoguriV}. Topological string amplitudes on each set of branes reduce to amplitudes in Chern-Simons theory, and the Ooguri-Vafa operator captures contributions from the scalar field describing strings stretched between these two sets of branes, whose amplitudes are identified with expectation values of Wilson loops in Chern-Simons theory \cite{Witten:1992fb,OoguriV}. According to the seminal work of Witten, such expectation values are identified with colored HOMFLY-PT polynomials \cite{Witten_Jones}, which are then assembled into a generating series that arises as the expectation value of the Ooguri-Vafa operator. The LMOV invariants that we consider are defined through factorization of this series \cite{OoguriV,Labastida:2000zp,Labastida:2000yw}. Upon the geometric transition, $N$ branes in the deformed conifold geometry are replaced by the resolved conifold geometry in the presence of additional lagrangian branes, which encode the topology of the original knot. Embedding this system in M-theory enables to interpret LMOV invariants as counting open M2-branes ending on M5-branes. However, integrality of these invariants has been verified only in some specific cases e.g. in \cite{OoguriV,Labastida:2000zp,Labastida:2000yw,Ramadevi_Sarkar,Mironov:2017hde}, as well as for some infinite families of knots and representations \cite{Garoufalidis:2015ewa,Kucharski:2016rlb}. In particular, in \cite{Kucharski:2016rlb} the relation of the framed unknot invariants (equivalently extremal invariants of twist knots, as well as open topological string amplitudes for branes in $\mathbb{C}^3$ geometry) to motivic Donaldson-Thomas invariants of the $m$-loop quiver was found, which led to the proof of integrality of BPS numbers in those cases; this relation was then analyzed and discussed also in \cite{Luo:2016oza,Zhu:2017lsn}.

\begin{figure}[b]
\begin{center}
\includegraphics[width=0.6\textwidth]{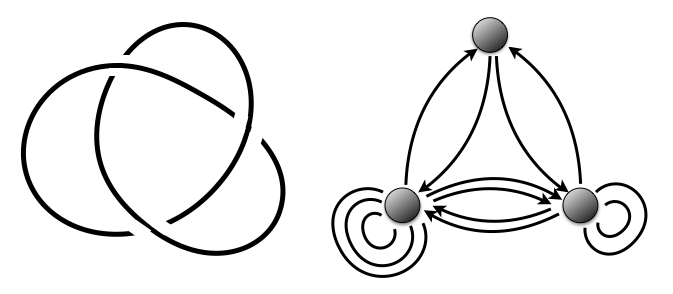} 
\caption{Trefoil knot and the corresponding quiver.}  \label{fig-trefoil}
\end{center}
\end{figure}

Reducing the above mentioned open M2-brane states to their worldvolume is expected to lead to a description in terms of $\mathcal{N}=4$ supersymmetric quiver quantum mechanics. We find this quantum mechanics description by postulating  that the Ooguri-Vafa generating function should be identified with the motivic generating series assigned to a putative quiver. Factorization of such a series defines motivic Donaldson-Thomas invariants, which also have an interpretation as the counts of BPS states \cite{Kontsevich:2008fj,Kontsevich:2010px}. If a quiver in question indeed exists, it is natural to identify these BPS states as the effective description of M2-M5 bound states in the Ooguri-Vafa description. As our main result -- announced already in \cite{Kucharski:2017poe} -- we show that the Ooguri-Vafa generating series indeed takes the form of the motivic generating series for some quiver, and we identify such quivers explicitly in various cases. For example, the quiver corresponding to the trefoil knot is shown in figure \ref{fig-trefoil}.

BPS states that arise in the quiver description can be interpreted as elements of Cohomological Hall Algebras \cite{Kontsevich:2010px,COM:8276935,Rei12}, which provide prototype examples of algebras of BPS states, whose existence was postulated in \cite{Harvey:1996gc}. These structures are intimately related to the theory of wall-crossing and associated phenomena, which led to important results both in physics and mathematics in recent years. In our work we take advantage of some of those results, as well as suggest new directions of studies. For example, it has been proved that motivic Donaldson-Thomas invariants assigned to a symmetric quiver are integer \cite{efimov2012}. Our results lead to the identification of LMOV invariants with motivic Donaldson-Thomas invariants for symmetric quivers, which thus proves integrality of these LMOV invariants. More precisely, for knots for which we identify the corresponding quiver, the LMOV conjecture for all symmetric representations is automatically proved.    
This is already an important result, and we expect that such corresponding quivers exist for all knots, and a general proof of the LMOV conjecture could be conducted along these lines. Some other identifications between quantities associated to knots and to quivers are shown in table \ref{tab-duality}. 

\begin{table}
\begin{center}
\begin{tabular}{c|c}
{\bf Knots} & {\bf Quivers} \\
\hline
Homological degrees, framing & Number of loops \\
Colored HOMFLY-PT & Motivic generating series \\
LMOV invariants & Motivic DT-invariants \\
Classical LMOV invariants & Numerical DT-invariants \\
Algebra of BPS states & Cohom. Hall Algebra \\
\end{tabular}
\caption{Identification of various quantities associated to knots and quivers.}  \label{tab-duality} 
\end{center}
\end{table}

There are many other consequences and new relations that follow from our work. First, motivic Donaldson-Thomas invariants that we consider have an interpretation as certain topological characteristics of quiver moduli spaces \cite{MR,FR}. This suggests that quiver moduli spaces themselves should be interpreted as knot invariants, which leads to a novel kind of categorification in knot theory. 

Second, we find that all HOMFLY-PT polynomials, as well as superpolynomials and their quadruply-graded generalizations, colored by arbitrary symmetric representations, are determined by a finite number of parameters: the matrix $C$ encoding the structure of the quiver corresponding to a given knot, and homological degrees of generators of the uncolored HOMFLY-PT homology. There should be a deeper reason why such limited information gives rise to rich structure and intricate properties of various infinite families of knot invariants.

Third, colored HOMFLY-PT polynomials and LMOV invariants can be defined for arbitrary (not only symmetric) representations and labeled by arbitrary Young diagrams. It is desirable to understand how this information is encoded in the corresponding quiver, or some generalization thereof. On the other hand, colored HOMFLY-PT polynomials labelled by symmetric representations satisfy a difference equation (encoded in $\widehat{A}$ operator), and their asymptotics is encoded in algebraic curves generalizing the A-polynomial \cite{AVqdef,Garoufalidis:2015ewa,Kucharski:2016rlb}. Such objects should also have an interesting interpretation in the context of quivers. In fact, for the $m$-loop quiver analogous functional equations have been discussed in \cite{COM:8276935}, and we expect that such relations should more generally play a role in quiver representation theory.

Fourth, having expressed colored HOMFLY-PT polynomials in the form of the motivic generating function, it is natural to replace one generating parameter associated with symmetric representations, by several parameters that naturally appear in motivic generating functions. This leads to a refinement of colored HOMFLY-PT polynomials, as well as LMOV invariants, and among others even stronger integrality statements.

Furthermore, motivic generating functions associated to quivers, as well as -- after our rewriting -- the generating functions of colored HOMFLY-PT polynomials, take the form of Nahm sums (with additional generating parameters) \cite{Nahm,Garoufalidis2015}. The Nahm sums have very intriguing properties, in special cases they are modular functions and may arise as characters of rational conformal field theories. It appears that both quiver representation theory, as well as knot theory, are rich sources of sums of this type.

Our work should also be related to many other results in literature. For example, uncolored HOMFLY-PT polynomials were related -- from a different perspective -- to Donaldson-Thomas invariants in \cite{Diaconescu-HOMFLY-DT}. A class of functions encoding integrality properties analogous to our generating functions has been analyzed in \cite{Schwarz:2013zua,Schwarz:2017gnp}. A detailed analysis of the LMOV conjecture was conducted in \cite{Dedushenko:2014nya}, and a refined LMOV conjecture was considered in \cite{Garoufalidis:2015ewa,Kameyama:2017ryw}.

This paper is organized as follows. In section \ref{sec-knots} we present appropriate background in knot theory and its relations to physics, including issues such as colored HOMFLY-PT polynomials, LMOV invariants, and knot homologies. In section \ref{sec-quivers} we introduce motivic Donaldson-Thomas invariants and other relevant notions from quiver representation theory. In section \ref{sec-knots-quivers} we present our main conjectures, motivated by physical interpretation of knot invariants in terms of supersymmetric quiver quantum mechanics, and relating various knot invariants to invariants of quivers. In section \ref{sec-results} we illustrate these conjectures in many examples, including infinite families of torus and twist knots, all knots with up to 6 crossings, and examples of thick knots. Using our results we also determine previously unknown HOMFLY-PT polynomials and superpolynomials colored by arbitrary symmetric representations for $6_2$ and $6_3$ knots, as well as for $(3,7)$ torus knot.


\section{Knot theory and physics}   \label{sec-knots}

Knot theory plays a prominent role in contemporary high energy and mathematical physics. As a branch of topology, it is not surprising that it is intimately related to topological field and string theories. It is perhaps more surprising, that through these links not only physics provides an interpretation of mathematical facts, but it is also a source of new ideas, which are subsequently formalized and (hoped to be) proved by mathematicians. Examples of such ideas, relevant in the context of our work, include Labastida-Mari{\~n}o-Ooguri-Vafa (LMOV) invariants, superplynomials and HOMFLY-PT homologies, quadruply-graded homologies, etc. In this section we recall and briefly summarize all these notions and introduce notation used in what follows.


\subsection{Knot invariants and the LMOV conjecture}

Polynomial knot invariants, including the Alexander polynomial known for almost 100 years, and the much younger Jones polynomial, form one important class of knot invariants. The Jones polynomial was subsequently generalized to the two-parameter HOMFLY-PT polynomial, and colored versions of these polynomials were introduced. Witten's interpretation of these polynomials as expectation values of Wilson loops in Chern-Simons theory \cite{Witten_Jones} played an important role in those developments. Furthermore, the Chern-Simons interpretation was also shown to be related to topological string theory \cite{Witten:1992fb}. This paved the way to subsequent formulation of LMOV invariants and the famous conjecture, stating that these invariants are integer \cite{OoguriV,Labastida:2000zp,Labastida:2000yw,Labastida:2001ts}. While this conjecture was verified in various specific situations
\cite{OoguriV,Labastida:2000zp,Labastida:2000yw,Ramadevi_Sarkar,Mironov:2017hde,Garoufalidis:2015ewa,Kucharski:2016rlb}, and some attempts of its general proof were undertaken \cite{Liu:2007kv}, it still appears to be an open problem. One aim of our work is to provide a proof of this conjecture, at least for a large class of knots and representations. However, let us first introduce relevant notation.

As follows from \cite{Witten_Jones}, colored HOMFLY-PT polynomials can be interpreted as expectation values of Wilson loops in representation $R$ in Chern-Simons theory 
\be
 \overline{P}_{R}(a,q) = \langle \textrm{Tr}_R U \rangle,
\ee
where $U=P\,\exp\oint_K A$ is the holonomy of $U(N)$ Chern-Simons gauge field along a knot $K$. Here the HOMFLY-PT polynomial is unreduced, i.e. it is normalized as
\be 
\overline{P}_{R}(a,q)=\overline{P}_R^{\bf 0_1} P_R(a,q),
\ee
where $P_R(a,q)$ is the corresponding reduced colored HOMFLY-PT polynomial (equal to 1 for the unknot), and $\overline{P}_R^{\bf 0_1}$ is the normalization factor of the unknot. As we will explain in what follows, our results depend in a crucial way on the choice of this normalization.

After embedding Chern-Simons theory in string theory, as we sketched in the introduction, it was shown in \cite{OoguriV} that the following generating function -- often referred to as the Ooguri-Vafa operator -- is natural to consider
\be
Z(U,V) = \sum_R  \textrm{Tr}_R U \, \textrm{Tr}_R V = \exp\Big(  \sum_{n=1}^{\infty} \frac{1}{n} \textrm{Tr} U^n \textrm{Tr} V^n \Big),
\ee
where $V$ is interpreted as a source, and the sum runs over all representations $R$, i.e. all two-dimensional partitions. The expectation value of this expression is the generating function of colored HOMFLY-PT polynomials. It was postulated in \cite{OoguriV,Labastida:2000zp,Labastida:2000yw,Labastida:2001ts} that this expectation value has the following structure
\be
\big\langle Z(U,V) \big\rangle = \sum_R \overline{P}_{R}(a,q) \textrm{Tr}_R V  = 
\exp \Big(  \sum_{n=1}^\infty \sum_R \frac{1}{n} f_{R}(a^n,q^n) \textrm{Tr}_R V^n  \Big),    \label{ZUV}
\ee
where the functions $f_R(a,q)$ take the form
\be
f_{R}(a,q) = \sum_{i,j} \frac{N_{R,i,j} a^i q^j}{q-q^{-1}}  \label{fR}
\ee
and encode conjecturally integer $N_{R,i,j}$ numbers. The functions $f_R(a,q)$ can be expressed as universal polynomials in colored HOMFLY-PT polynomials. The above statements, concerning the structure of $\big\langle Z(U,V) \big\rangle$ and integrality of $N_{R,i,j}$, are referred to as the LMOV conjecture, and $N_{R,i,j}$ are called LMOV invariants (or Ooguri-Vafa invariants). As indicated in the introduction, in the physics interpretation they count bound states of M2-branes ending on M5-branes.

Of our main interest in this work is the generating function of $S^r$-colored HOMFLY-PT polynomials. It can be obtained by considering a one-dimensional source $V=x$. In this case $\textrm{Tr}_R V \neq 0$ only for symmetric representations $R=S^r$, and then $\textrm{Tr}_{S^r}(x) = x^r$. Then (\ref{ZUV}) reduces to the generating function of $S^r$-colored HOMFLY-PT polynomials, and denoting the $S^r$-colored HOMFLY-PT polynomial by $\overline{P}_{r}(a,q)$ we get
\be
P(x) = \langle Z(U,x) \rangle = \sum_{r=0}^\infty \overline{P}_{r}(a,q) x^r =   
\exp\Big({\sum_{r,n\geq 1} \frac{1}{n} f_{r}(a^n,q^n)x^{n r}}\Big),
\label{Pz2}
\ee
with  
\be
f_r(a,q)\equiv f_{S^r}(a,q) = \sum_{i,j} \frac{N_{r,i,j} a^i q^j}{q-q^{-1}},
\ee
where LMOV invariants are denoted by $N_{r,i,j}\equiv N_{S^r,i,j}$. These functions are polynomials, with rational coefficients, of $\overline{P}_{d_1}(a^{d_2},q^{d_2})$ for  some $d_1$ and $d_2$ (with $d_1 d_2\le r$): 
\begin{align}
f_1(a,q) &= \overline{P}_1(a,q), \nonumber\\
f_2(a,q) &=  \overline{P}_2(a,q) - \frac{1}{2} \overline{P}_1(a,q)^2 -\frac{1}{2}  \overline{P}_1(a^2,q^2), \nonumber \\
f_3(a,q) &= \overline{P}_3(a,q) - \overline{P}_1(a,q)\overline{P}_2(a,q) + \frac{1}{3}\overline{P}_1(a,q)^3 - \frac{1}{3} \overline{P}_1(a^3,q^3), \nonumber
\end{align}
etc. One can also rewrite (\ref{Pz2}) in the product form
\be
P(x) = \prod_{r\geq 1;i,j;k\geq 0} \Big(1 - x^r a^i q^{j+2k+1} \Big)^{N_{r,i,j}}.
\label{Pr-LMOV} 
\ee
One of our aims is to show integrality of BPS degeneracies $N_{r,i,j}$ encoded in this product.

In the (classical) limit $q\to 1$, a special role is played by a subset of LMOV invariants, referred to as classical LMOV invariants. To define them it is useful to consider the ratio 
\be
y(x,a) =  \lim_{q\to 1} \frac{P(qx)}{P(x)} = 
\lim_{q\to 1}  \prod_{r\geq 1;i,j;k\geq 0} \Big(\frac{1 - x^r a^i q^{r+j+2k+1} }{1 - x^r a^i q^{j+2k+1}}\Big)^{N_{r,i,j}}  
= \prod_{r\geq 1;i} (1 - x^r a^i)^{-r b_{r,i} / 2},  \label{yxa}
\ee
with classical LMOV invariants defined as
\be
b_{r,i}  = \sum_j N_{r,i,j}.   \label{bri}
\ee
It turns out that $y=y(x,a)$ defined above satisfy algebraic equations 
\be
A(x,y)=0
\ee
of A-polynomial type \cite{AVqdef,Garoufalidis:2015ewa}.


\subsection{Knot homologies}    \label{ssec-knothom}

Another important class of knot invariants are knot homologies. First well understood examples of such structures are Khovanov homology \cite{Khovanov} and Khovanov-Rozansky homology \cite{KhR1,KhR2}. In our work an important role is played by their putative, highly nontrivial generalization, namely colored HOMFLY-PT homology $\mathcal{H}^{S^r}_{i,j,k}$, which categorifies the HOMFLY-PT polynomial colored by symmetric representations $S^r$. It has been defined rigorously by mathematicians only recently \cite{Cautis}, yet only for the unreduced version, and it is still not suitable for explicit computations (there also exist some constructions in the case of antisymmetric representations $\Lambda^r$, both reduced and unreduced versions, see e.g. \cite{Wedrich:2016smm}, which are conjecturally isomorphic to the homologies corresponding to the symmetric representations). Nonetheless, the conjectural Poincar\'{e} polynomial of (reduced) colored HOMFLY-PT homology, referred to as the superpolynomial
\be
P_r (a,q,t) = \sum_{i,j,k} \, a^i q^j t^k \, \dim \mathcal{H}^{S^r}_{i,j,k},    \label{superpolynomial}
\ee
can be determined for various families of knots, for example using the formalism of differentials and the structural properties of the (colored) homologies \cite{DGR,Rasmussen-differentials,Gukov:2011ry,Gukov:2015gmm} -- the formalism that we will exploit in the present paper. It has been postulated that HOMFLY-PT homology should be identified with the space of BPS states in relevant brane systems \cite{GSV,Gukov:2011ry}. Superpolynomials for the unknot or the Hopf-link can be also computed by techniques of refined topological string theory \cite{GIKV}, and superpolynomials for torus knots can be computed by means of refined Chern-Simons theory \cite{AS,FGS,DMMSS,superA,ItoyamaMMM}. Colored superpolynomials reduce to colored HOMFLY-PT polynomials upon the substitution $t=-1$. As we will see, one interesting result of our work is an explicit relation between colored HOMFLY-PT polynomials and the (uncolored) superpolynomial.

Let us briefly present structural properties of the reduced $S^r$-colored HOMFLY-PT homologies $\mathcal{H}^{S^r}$ of a given knot \cite{Gukov:2011ry,Gorsky:2013jxa}. First, for a given knot, for every $k=0,\ldots,r-1$, there exists a (positive, vertical) colored differential $d_{1-k}$ on $\mathcal{H}^{S^r}$, of $(a,q,t)$-degree $(-2,2-2k,-1)$, such that the homology of $\mathcal{H}^{S^r}$ with respect to $d_{1-k}$ is isomorphic to $\mathcal{H}^{S^k}$. Second, for every $k=0,\ldots,r-1$ there is another set of (negative, vertical) colored differentials $d_{-r-k}$ on $\mathcal{H}^{S^r}(K)$, of $(a,q,t)$-degree $(-2,-2r-2k,-3)$,
such that the homology of $\mathcal{H}^{S^r}(K)$ with respect to $d_{-r-k}$ is isomorphic to $\mathcal{H}^{S^k}$. Third, there is a universal colored differential $d_{2\to 1}$
of degree $(0,2,0)$ on the homology $\mathcal{H}^{S^2}(K)$, such that the homology of $\mathcal{H}^{S^2}$ with respect to $d_{2\to 1}$ is isomorphic to the uncolored homology $\mathcal{H}^{S}$. All these differentials relate homology theories with different values of $r$. The uncolored homology, corresponding to $r=1$, supposedly categorifies the reduced HOMFLY-PT polynomial, and its Poincar\'e polynomial is simply the original (uncolored) superpolynomial introduced in \cite{DGR}. 

Furthermore, HOMFLY-PT homologies of a large class of knots satisfy the refined exponential growth, which implies the following relation for their colored superpolynomials 
\be
P_{S^r}(a,q=1,t)=\left( P_{\tableau{1}}(a,q=1,t) \right)^r,   \label{exp-growth}
\ee
see also \cite{Wedrich:2016smm}. Properties of colored differentials, together with the assumption of the exponential growth, enable to determine an explicit form of the colored superpolynomial $P_{S^r}(a,q,t)$ for various knots \cite{Gukov:2011ry,superA,FGSS}. For example, colored superpolynomials for the trefoil knot $3_1$ take the form \cite{superA,FGSS}
\be
P_r(a,q,t)=  \frac{a^{2r}}{q^{2r}}\sum_{k=0}^r {r \brack k} q^{2k(r+1)} t^{2k} \prod_{i=1}^k  (1+a^2q^{2(i-2)}t), \label{Pr-31}
\ee
where
\be
{r \brack k} = \frac{(q^2;q^2)_r}{(q^2;q^2)_k(q^2;q^2)_{r-k}},   \label{q-binomial}
\ee
and the $q$-Pochhammer symbol is defined as
\be
(x;q)_r =\prod_{k=0}^{r-1} (1-x q^k),\qquad \qquad   (x;q)_\infty =\prod_{k=0}^\infty (1-x q^k) \,.    \label{qdilog}
\ee
For $t=-1$, (\ref{Pr-31}) specializes to the reduced colored HOMFLY-PT polynomial for trefoil knot, while in the uncolored case (i.e. for $r=1$), (\ref{Pr-31}) reduces to
\be
P_1(a,q,t) = \frac{a^2}{q^2} + a^2 q^2 t^2 + a^4 t^3.   \label{P1-31}
\ee
The monomials in this expression correspond to generators of the HOMFLY-PT homology, and powers of $t$ in each monomial -- in this example taking values $(0,2,3)$ -- are referred to as homological degrees.

\begin{figure}[b]
\begin{center}
\begin{tikzpicture}
\draw [help lines] (0,0) grid (3,2);
\draw [very thick] (0,2.2)--(0,0)--(3.2,0);
\draw [thick] (-1,-1)--(0,0);
\draw [thick] (0,-1)--(0,0);
\draw [thick] (-1,0)--(0,0);
\draw (-0.7,-0.3) node {{\textbf{\it a}}};
\draw (-0.3,-0.7) node {{\it q}};
\filldraw [black] (0.5,0.5) circle  (2pt);
\filldraw [black] (1.5,1.5) circle  (2pt);
\filldraw [black] (2.5,0.5) circle  (2pt);
\draw (0.33,0.72) node {$0$};
\draw (1.3,1.7) node {$3$};
\draw (2.65,0.7) node {$2$};
\draw (-0.45,0.5) node {$2$};
\draw (-0.45,1.5) node {$4$};
\draw (0.48,-0.5) node {$-2$};
\draw (1.55,-0.5) node {$0$};
\draw (2.55,-0.5) node {$2$};
\draw [->,>=stealth,thin] (1.6,1.4)--(2.4,0.6);
\draw [->,>=stealth,thin] (1.4,1.4)--(0.6,0.6);

\draw [help lines] (7,-1) grid (10,2);
\draw [very thick] (7,2.2)--(7,-1)--(10.2,-1);
\draw [thick] (7,-1)--(6,-2);
\draw [thick] (7,-1)--(6,-1);
\draw [thick] (7,-1)--(7,-2);
\draw (6.3,-1.3) node {{\textbf{\it a}}};
\draw (6.7,-1.7) node {{\it q}};
\filldraw [black] (7.5,0.5) circle  (2pt);
\filldraw [black] (8.5,0.5) circle  (2pt);
\filldraw [black] (9.5,0.5) circle  (2pt);
\filldraw [black] (8.5,1.5) circle  (2pt);
\filldraw [black] (8.5,-0.5) circle  (2pt);
\draw (7.38,-1.5) node {$-2$};
\draw (8.45,-1.5) node {$0$};
\draw (9.45,-1.5) node {$2$};
\draw (6.45,1.5) node {$2$};
\draw (6.45,0.5) node {$0$};
\draw (6.39,-0.5) node {$-2$};
\draw (7.32,0.73) node {$-1$};
\draw (8.5,0.73) node {$0$};
\draw (9.63,0.73) node {$1$};
\draw (8.4,1.75) node {$2$};
\draw (8.3,-0.77) node {$-2$};
\draw [->,>=stealth,thin] (8.6,1.4)--(9.4,0.6);
\draw [->,>=stealth,thin] (8.4,1.4)--(7.6,0.6);
\draw [->,>=stealth,thin] (7.6,0.4)--(8.4,-0.4);
\draw [->,>=stealth,thin] (9.4,0.4)--(8.6,-0.4);
\end{tikzpicture}
\end{center}
\caption{Diagrams of the reduced uncolored HOMFLY-PT homology of the trefoil (left) and the figure-eight knot (right).}
\label{diagrams}
\end{figure}
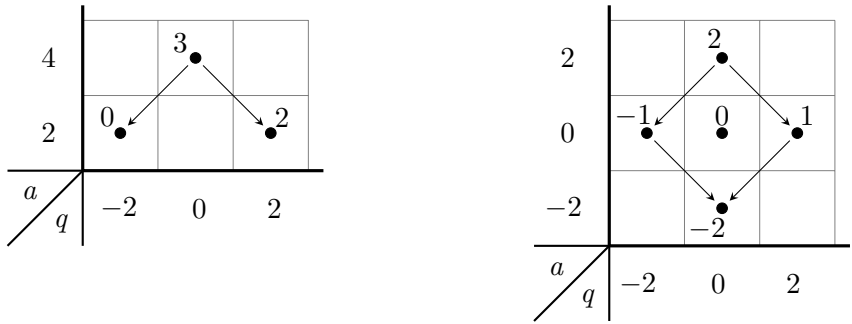


It is convenient to present the structure of (colored) HOMFLY-PT homology in terms of diagrams on a two-dimensional lattice. Each homology generator is represented by a dot at position $(i,j)$ in such a lattice, with $i$ and $j$ representing respectively $q$-degree and $a$-degree of this generator, and whose $t$-degree can in addition be written explicitly in the diagram as label of a corresponding dot. In addition, differentials acting between pairs of generators can be represented by arrows in the diagram. In figure \ref{diagrams} are presented the diagrams of the reduced uncolored HOMFLY-PT homologies of the trefoil and the figure-eight knot.

The structure of differentials implies that in the case of the uncolored homologies (the ones that are of our main interest in the paper) the generators must form two larger structures, which we call a zig-zag and a diamond. A zig-zag is a string of an odd number of generators, and a diamond consists of four generators (as the name indicates, distributed in the form of a diamond). For each knot, its HOMFLY-PT homology must contain precisely one zig-zag (possibly of length one, i.e. consisting of a single generator), and an arbitrary number of diamonds. For example, in the case of the diagrams in figure \ref{diagrams}, the diagram of the trefoil knot consists of a single zig-zag of length three, while the diagram for the figure-eight knot consists of a zig-zag of length one (only the generator in the middle of the diagram with the label 0), and one diamond formed by the remaining four generators. The (finite-dimensional) homology of a link has as many zig-zags as the number of its components. 

The structure of colored HOMFLY-PT homology was further generalized to quadruply-graded homology, which has a richer structure of differentials \cite{Gorsky:2013jxa}. The Poincar\'e polynomial of this quadruply-graded homology $P_r(a,Q,t_r,t_c)$ depends on four parameters $a,Q,t_r$ and $t_c$, and specializes to the colored superpolynomial upon the identification
\be
P_r(a,q,t) = P_r(a,Q=q,t_r=t q^{-1},t_c=q),    \label{quad-super}
\ee
and to the colored HOMFLY-PT polynomial upon
\be
P_r(a,q) = P_r(a,Q=q,t_r=-q^{-1},t_c=q).  \label{super-homfly}
\ee
Quadruply-graded homologies for a large class of knots satisfy the refined exponential growth, which implies the following relation for the corresponding Poincar\'e polynomials 
\be
P_r(a,Q,t_r,t_c=1)=\big(P_1(a,Q,t_r,t_c=1)\big)^r.   \label{exp-growth-quad}
\ee



\section{Quiver moduli and Donaldson-Thomas invariants}   \label{sec-quivers}

In this section we present basic properties of quivers and moduli spaces of their representations, which will be crucial in the rest of the paper. Moduli spaces of quiver representations have a rich structure, which among others provides a natural playground for the theory of (motivic) Donaldson-Thomas invariants, Cohomological Hall Algebras, etc. 

A quiver $Q$ is an oriented graph with a finite set of vertices $Q_0$ and finitely many arrows between vertices $\alpha:i\rightarrow j$, for $i,j \in Q_0$. On $\mathbb{Z}Q_0$, we define the Euler form of $Q$ by
\be
\langle {\bm d},{\bm e}\rangle_Q=\sum_{i\in Q_0}d_ie_i-\sum_{\alpha:i\rightarrow j}d_ie_j.
\ee
A quiver representation assigns to each vertex $i\in Q_0$ a vector space of dimension $d_i$, and a linear map between two such spaces to each arrow. The vector ${\bf d}=(d_1,\ldots,d_m)$ is referred to as the dimension vector.

As we will see, quivers which appear in relation to knot invariants are symmetric, meaning that for any pair of their vertices $i$ and $j$, the number of arrows from $i$ to $j$ is equal to the number of arrows from $j$ to $i$. While explicit expressions for invariants describing moduli spaces of quiver representations are hard to find in general, they are quite well understood in the case of symmetric quivers \cite{Kontsevich:2010px,efimov2012,FR,MR}. An important information about the moduli space of representations of a symmetric quiver is encoded in the following generating series 
\be
P_Q({\bm x})=\sum_{{\bm d}\in\mathbb{N}Q_0}(-q)^{-\langle {\bm d},{\bm d}\rangle_Q} {\bm x}^{\bm d} \prod_{i\in Q_0}\prod_{j=1}^{d_i}\frac{1}{1-q^{-2j}}   \label{PQx}
\ee
where ${\bm x}^{\bm d}=\prod_{i\in Q_0}x_i^{d_i}$. In particular, motivic Donaldson-Thomas invariants $\Omega_{{\bm d},j}\equiv \Omega_{d_1,\ldots,d_m;j}$, assembled into 
\be\label{OmegaPQx}
\Omega_{\bm d}(q) = \sum_j \Omega_{{\bm d},j} (-q)^j,
\ee
are defined once (\ref{PQx}) is rewritten as
\be
P_Q({\bm x})={\rm Exp}(\frac{1}{q^{-1}-q}\sum_{{\bm d}\neq0}\Omega_{\bm d}(q){\bm x}^{\bm d}),
\ee
where ${\rm Exp}$ is the plethystic exponential defined by ${\rm Exp}(f+g)={\rm Exp}(f)\cdot{\rm Exp}(g)$ and ${\rm Exp}(q^ix^d)=\frac{1}{1-q^ix^d}$ for $i\in\mathbb{Z}$ and $d\in\mathbb{N}Q_0$. This plethystic exponential form can be written equivalently as a product decomposition
\be
P_Q({\bm x})=
\prod_{{\bm d}\neq 0} \prod_{j\in\mathbb{Z}} \prod_{k\geq 0} \big(1 - (-1)^j {\bm x}^{\bm d} q^{j+2k+1} \big)^{-\Omega_{{\bm d},j}}.   \label{PQx-Omega}
\ee
Two geometric interpretations of invariants $\Omega_{\bm d}(q)$, either as the intersection Betti numbers of the moduli space of all semisimple representations of $Q$ of dimension vector ${\bm d}$, or as the Chow-Betti numbers of the moduli space of all simple representations of $Q$ of dimension vector ${\bm d}$, were provided in \cite{MR,FR}. It was also proved these invariants are positive integers \cite{efimov2012}. One of our aims in this paper is to relate these invariants to LMOV invariants of knots.

One can also introduce numerical Donaldson-Thomas invariants of a quiver. To this end, for a vector ${\bm n}\in\mathbb{N}Q_0$, we denote by $P_Q(q^{\bm n}{\bm x})$ the series arising from $P_Q({\bm x})$ by replacing every ${\bm x}^{\bm d}$ by $q^{{\bm n}\cdot {\bm d}} {\bm x}^{\bm d}$, where ${\bm n}\cdot {\bm d}=\sum_{i\in Q_0}n_id_i$. Then
\be
\frac{P_Q((-q)^{\bm n}{\bm x})}{P_Q((-q)^{-\bm n}{\bm x})}={\rm Exp}(\sum_{d\not=0}\frac{(-q)^{{\bm n}\cdot {\bm d}}-(-q)^{-{\bm n}\cdot {\bm d}}}{(-q)-(-q)^{-1}}\Omega_{\bm d}(q){\bm x}^{\bm d}),
\ee
and in this equation we can specialize $q$ to $1$. By \cite{Engel2009}, the left hand side specializes then to the generating series of the Euler characteristic of certain Hilbert schemes ${\rm Hilb}_{d,n}(Q)$ attached to the quiver (these numbers admit a combinatorial interpretation by counting certain kinds of trees), so that we get
\be
\sum_{d\in\mathbb{N}Q_0}\chi({\rm Hilb}_{{\bm d},{\bm n}}(Q)){\bm x}^{\bm d}={\rm Exp}(\sum_{{\bm d}\not=0}({\bm n}\cdot {\bm d})\Omega_d(1) {\bm x}^{\bm d})=\prod_{{\bm d}\not=0}(1-{\bm x}^{\bm d})^{-({\bm n}\cdot {\bm d})\Omega_{\bm d}(1)}.   \label{Omega-classical}
\ee
These $\Omega_{\bm d}(1)$ are the (numerical) Donaldson-Thomas invariants of the quiver.



\section{Knot invariants from quivers}   \label{sec-knots-quivers}

In this section we first present our main claim, relating various knot invariants to quivers. We also discuss its  various implications, and develop a formalism facilitating computations and enabling to determine quivers associated to knots. Our claim takes form of the following conjectures. We show that these conjectures are correct in many explicit and nontrivial examples in section \ref{sec-results}.


\subsection{Main conjectures}

\begin{conjecture}\label{PxCConj}

For a given knot, the generating function of its (appropriately normalized, as explained in detail in section \ref{ssec-unreduced}) colored HOMFLY-PT polynomials (\ref{Pz2}) can be written in the form
\be
P(x)  = \sum_{r=0}^{\infty} \overline{P}_r(a,q) x^r = \sum_{d_1,\ldots,d_m\geq 0} x^{d_1+\ldots + d_m}  q^{\sum_{i,j} C_{i,j} d_i d_j}    \frac{\prod_{i=1}^m q^{l_i d_i} a^{a_i d_i}(-1)^{t_i d_i}}{\prod_{i=1}^m(q^2;q^2)_{d_i}}  \label{PxC}
\ee
where $C$ is a (symmetric) $m\times m$ matrix, and $l_i,a_i$ and $t_i$ are fixed integers. Note that terms proportional to $x^r$, with fixed $r$, arise from sets of $\{d_i \}$ such that $r=d_1+\ldots +d_m$. Remarkably, (\ref{PxC}) has the same form as the motivic generating function (\ref{PQx}) of a symmetric quiver determined by the matrix $C$, up to the identification $q \mapsto -q$ and the specialization of variables
\be
x_i = x a^{a_i} q^{l_i-1} (-1)^{t_i}.     \label{specialize}
\ee
The number of vertices $m$ of such a quiver is given by the size of $C$, and the number of arrows between vertices $i$ and $j$ is given by the matrix element $C_{i,j}$ (in particular $C_{i,i}$ denotes the number of loops at vertex $i$).

\end{conjecture}

It follows that to a given knot one can assign a quiver, so that various invariants of this knot are encoded in the data of moduli spaces of quiver representations of this corresponding quiver. Moreover, all this information is encoded in a finite set of parameters that determine (\ref{PxC}): the matrix $C$, as well as integers $l_i,a_i,t_i$ that are encoded in the (uncolored, reduced) superpolynomial of the knot in question. Recall that the uncolored, reduced superpolynomial for a given knot is a sum of monomials of the form $a^{a_i} q^{q_i} t^{t_i}$, which correspond to generators of the HOMFLY-PT homology. 

\begin{conjecture}  \label{conj2}
The size of the matrix $C$ (the number of vertices in the corresponding quiver) is equal to the number of generators of uncolored HOMFLY-PT homology. Furthermore, with appropriate ordering of vertices, $t_i$ in (\ref{PxC}) agree with homological degrees of generators of HOMFLY-PT homology, diagonal elements of $C$ are also equal to homological degrees, i.e. $C_{i,i}=t_i$, coefficients of linear powers of $q$ take the form $l_i=q_i-t_i$, and $a_i$ are equal to $a$-degrees of generators of uncolored HOMFLY-PT homology. An additional minus sign in (\ref{PxC}) comes with the power determined by $t_i$, so that it is relevant only for  generators with odd $t$-grading. 
\end{conjecture}

Note that it follows that homological degrees $t_i$ can be identified (as diagonal elements of matrix $C$) after rewriting the generating series (\ref{Pz2}) in the quiver-like form, and they are given by the number of loops in the corresponding quiver. This means, that the uncolored superpolynomial is encoded in the form of colored HOMFLY-PT polynomials, which is quite a surprising observation.

For knots that satisfy the refined exponential growth (\ref{exp-growth}) it is not hard to see where the coefficients of linear terms in $d_i$, in powers of $q,a$ and $(-1)$ in (\ref{PxC}), come from. In general, in expressions for quadruply-graded superpolynomials, $t_c$ does not appear in powers which are linear in summation variables, so -- if (\ref{exp-growth-quad}) holds -- linear powers of other parameters can be identified upon specialization $t_c=1$. Furthermore, recall that colored superpolynomials arise upon the substitution (\ref{quad-super}), and colored HOMFLY-PT polynomials upon (\ref{super-homfly}). It follows that for arbitrary color $r=d_1+\ldots+d_m$, a linear term in the exponent of $q$ takes the form
\begin{equation}
\sum_i (q_i-t_i) d_i,    \label{linear}
\end{equation}
where the sum is over all generators $i$ of the uncolored homology, $q_i$ and $t_i$ are their $q$- and $t$-degrees, and $d_i$ is the corresponding summation index. Analogously, linear powers of parameters $a$ and $(-1)$ in (\ref{PxC}) must, respectively, take the form $\sum_i a_i d_i$ and $\sum_i t_i d_i$. The same formulas are valid for the unreduced homology, since the refined exponential growth holds for the unreduced homology of the unknot.

Note that one can also focus on those parts of colored HOMFLY-PT polynomials or superpolynomials which are proportional to the  highest or lowest power of the variable $a$ \cite{Gorsky:2013jxa,Garoufalidis:2015ewa}. The corresponding generators of HOMFLY-PT homology lie respectively in the top or bottom row of the homology diagram, so such invariants are often referred to as top/bottom row invariants, or extremal invariants. For a large class of knots satisfying the exponential growth property (\ref{exp-growth}), the generating function of their colored extremal reduced HOMFLY-PT polynomials also takes a universal form (\ref{PxC}), however with the dependence on $a$ suppressed
\begin{equation}
P^{bottom/top}(x)= \sum_{d_1+d_2+\cdots+d_m\geq 0} x^{d_1+\ldots +d_m} q^{\sum_{i,j} C_{i,j} d_i d_j}\frac{q^{\sum_i (q_i-t_i) d_i} (-1)^{\sum_i {t_i d_i} }}{\prod_{i=1}^m (q^2;q^2)_{d_i}}.     \label{quivgen2}
\end{equation}
Here $m$ is the dimension of the fundamental homology corresponding to the bottom/top row, and $q_i$ and $t_i$, $i=1,\ldots,m$, are $q$-degrees and $t$-degrees of these $m$ generators. In this case the matrix $C$ encodes a quiver which is a subquiver (capturing only extremal $a$-dependence) of the full quiver associated to a given knot.

The conjecture \ref{conj2} relates various quantities associated to knots to those of quiver moduli. Note that other relations of this type also follow -- one another example of such a relation is the dependence on framing. The operation of framing by $f\in \mathbb{Z}$ changes the colored HOMFLY-PT polynomial by a factor, which for the symmetric representation $S^r$ takes the form 
\be
a^{2fr}q^{f \, r(r-1)}.    \label{framing}
\ee
From the viewpoint of the quiver generating function (\ref{PxC}), the term with quadratic (in $r$) power of $q$
\be
q^{fr^2}=q^{f(\sum_i d_i)^2} = q^{f\sum_{i,j} d_i d_j}
\ee
shifts all elements of the matrix $C$ by $f$
\be
C\ \mapsto \ C+f\left[\begin{array}{cccc} 1&1&\cdots&1\\
1&1&\cdots&1\\
\vdots&\vdots&\ddots&\vdots\\
1&1&\cdots&1
\end{array}\right]     \label{C-frame}
\ee
which in the quiver interpretation corresponds to adding $f$ loops at each vertex and $f$ pairs of oppositely-oriented arrows between all pairs of vertices. Note that, while in the context of quivers all entries of a matrix $C$ should be nonnegative, in some examples coming from knots we find matrices $C$ with negative entries. In this case a change of framing can be used to shift such negative values, and make all entries of $C$ nonnegative; such a modified matrix still describes the same knot.

Furthermore, we can also characterize the structure of the matrix $C$ in more detail.

\begin{conjecture}   \label{conj3}
For a given knot, the matrix $C$ has a block structure
$$
C = \left[
\begin{array}{c|c|c|c}
b_{1,1} & \cdots & b_{1,k} & \cdots \\ \hline 
 \vdots & \ddots & \vdots &  \\ \hline
b_{1,k}^T & \cdots & b_{k,k} & \\ \hline
\vdots& & & \ddots   \\
\end{array}
\right]
$$
Diagonal blocks $b_{k,k}$ correspond to structural elements of the HOMFLY-PT homology, introduced in section \ref{ssec-knothom}. One of those blocks (in case of knots; or as many blocks as the number of components of a link) corresponds to the zig-zag element of length $2p+1$, and it has the same form (up to some permutation of homology generators, and up to an overall shift by a constant matrix with integer coefficients as in (\ref{C-frame}), corresponding to framing) as the matrix $C$ for the $(2,2p+1)$ torus knot (\ref{C-torus22p1}). All other diagonal block elements correspond to diamonds and (up to a permutation of homology generators) take the form
\be
\left[\begin{array}{cccc} k&k&k+1&k+1\\
k&k+1&k+2&k+2\\
k+1&k+2&k+3&k+3\\
k+1&k+2&k+3&k+4
\end{array}\right]    \label{diamond}
\ee 
for a fixed $k$ (which may be different for each block). The structure of other blocks $b_{l,k}$ for $l\neq k$ depends only on the structural elements corresponding to diagonal blocks $b_{l,l}$ and $b_{k,k}$.

\end{conjecture}


\subsection{Superpolynomials and quadruply-graded homology of knots from quivers}

In the above conjectures we related generating functions of colored HOMFLY-PT polynomials to the motivic generating series of some quiver. However, in addition we postulate that the same quiver encodes also generating functions of colored superpolynomials, as well as Poincar{\'e} polynomials of quadruply-graded HOMFLY-PT homology.

\begin{conjecture}\label{PxCConj-quadruply}

Consider a knot satisfying the exponential growth property (\ref{exp-growth-quad}), with the corresponding quiver -- determined as explained above -- represented by a matrix $C$, the size of the reduced colored homology denoted by $m$, and $(a,q,t)$-degrees of its generators denoted by $(a_i,q_i,t_i)$. Then, the Poincar{\'e} polynomial of the reduced quadruply-graded $S^r$-colored homology is also determined by the quiver matrix $C$, and it takes a universal form
\begin{align}
P_r(a,Q,t_r,t_c)=& \sum_{d_1+d_2+\ldots+d_m=r}  \frac{(t_c^2;t_c^2)_r}{(t_c^2;t_c^2)_{d_1}(t_c^2;t_c^2)_{d_2} \cdots (t_c^2;t_c^2)_{d_m}} \times  \nonumber \\
& \qquad \times  a^{\sum_{i=1}^m a_i d_i}Q^{\sum_{i=1}^m q_i d_i} t_r^{\sum_{i=1}^m t_i d_i} t_c^{\sum_{i,j=1}^m {C_{i,j} d_i d_j} } .   \label{quad-C}
\end{align}

\end{conjecture}

The generating function of such Poincar{\'e} polynomials, normalized by $(t_c^2;t_c^2)_r$, 
\be
P(x) = \sum_{r=0}^{\infty} P_r(a,Q,t_r,t_c) \frac{x^r}{(t_c^2;t_c^2)_r},
\ee
can also be obtained as a specialization of (\ref{PQx}), with appropriate choice of $x_i$, which then gives rise to linear (in $d_i$) powers of $a,Q$ and $t_r$. Therefore the product decomposition (\ref{PQx-Omega}) leads to refined (quadruply-graded) LMOV invariants. Furthermore, (\ref{quad-C}) can be reduced to the generating function of colored superpolynomials upon the identification of variables given in (\ref{quad-super}), and the corresponding refined LMOV invariants can be identified (note that LMOV invariants, refined in the sense of including $t$-dependence, were also discussed in \cite{Garoufalidis:2015ewa,Kameyama:2017ryw}). 

The expression (\ref{quad-C}) can also be reduced to the case of extremal powers of $a$ (i.e. top/bottom row). Let now $m$ denote the size of such a bottom or top row uncolored reduced homology, and denote $(q,t)$-degrees of its generators by $(q_i,t_i$). The Poincar{\'e} polynomial of the bottom (or top) row of the reduced quadruply-graded $S^r$-colored homology is then given by
\begin{equation}
P^{bottom/top}_r(Q,t_r,t_c)=\sum_{d_1+d_2+\ldots+d_m=r}  Q^{\sum_{i=1}^m q_i d_i} t_r^{\sum_{i=1}^m t_i d_i} t_c^{\sum_{i,j=1}^m {C_{i,j} d_i d_j} } \frac{(t_c^2;t_c^2)_r}{(t_c^2;t_c^2)_{d_1} \cdots (t_c^2;t_c^2)_{d_m}}.   \label{quad-extreme}
\end{equation}
In this case we suppressed the $a$-dependence, since the entire bottom (or top) row homology is characterized by the same $a$-degree. The matrix $C$ in (\ref{quad-extreme}) coincides with the one in (\ref{quivgen2}) and it encodes a subquiver (representing only the extremal $a$-dependence) of the full quiver associated to a given knot. Specializing the product decomposition (\ref{PQx-Omega}) results in refined (quadruply-graded), extremal LMOV invariants. Moreover (\ref{quad-extreme}) can be reduced to the generating function of (extremal) colored superpolynomials upon the identification of variables given in (\ref{quad-super}), which then encodes refined extremal LMOV invariants.

We stress that integrality of various refined (or quadruply-graded) LMOV invariants mentioned above follows automatically from the fact that the corresponding generating series arise as specializations of (\ref{PQx-Omega}), whose product decomposition is proved to give rise to integer invariants in general.



\subsection{Consequences: proof of the LMOV conjecture, new categorification, etc.}

Our conjectures imply that various knot invariants can be expressed in terms of invariants characterizing quiver moduli spaces. More generally, these conjectures imply that there are various -- unexpected, and highly nontrivial -- relations between knot theory and quiver representation theory; some of those relations are listed in table \ref{tab-duality}. We now briefly discuss some of these consequences, and we will illustrate them in various examples in the next section. 

First, the fact that -- under appropriate specialization -- the motivic generating series of a quiver agrees with the generating function of colored HOMFLY-PT polynomials also means, that the product decomposition (\ref{PQx-Omega}) is identified with the product decomposition (\ref{Pr-LMOV}). This implies that the LMOV invariants $N_{r,i,j}$ take the form of linear combinations (with integer coefficients) of motivic Donaldson-Thomas invariants $\Omega_{{\bm d},j} = \Omega_{d_1,\ldots,d_m;j}$. The motivic Donaldson-Thomas invariants for symmetric quivers are proved to be integer \cite{efimov2012}, which therefore implies that the corresponding LMOV invariants are also integer -- which then proves the LMOV conjecture. Therefore, once a quiver corresponding to a given knot is identified (which we will do in many examples in the rest of the paper), it automatically follows that LMOV invariants for this knot, labeled by symmetric representations, are integer. 

Second, quiver invariants automatically provide a refinement of knot invariants -- once a quiver is identified, its motivic generating series (\ref{PQx}) involves several generating parameters $x_1,\ldots,x_m$, encoding ``refined'' invariants $\Omega_{d_1,\ldots,d_m;j}$, and ``refined'' HOMFLY-PT polynomials. It is desirable to understand the meaning of those refined invariants from the knot theory perspective.

Third, we find that in some cases to a given knot one may assign several quivers, which give rise to the same generating function of HOMFLY-PT polynomials -- even though their original motivic generating series, without imposing the specialization (\ref{specialize}), are different. Such quivers differ by some permutation of their elements, as we will illustrate in various examples. 

Moreover, the limit $q\to 1$ of the motivic generating series immediately implies integrality of classical LMOV invariants (\ref{bri}), which are expressed in terms of (integer) numerical Donaldson-Thomas invariants defined in (\ref{Omega-classical}). 

Furthermore, the fact that (generating functions of) colored HOMFLY-PT polynomials and LMOV invariants are expressed in terms of motivic Donaldson-Thomas invariants -- which arise as certain Betti numbers of quiver moduli spaces -- provides a novel categorification of these knot invariants. Namely, quiver moduli spaces themselves can be regarded as new invariants of knots. 

While, on one hand, knot invariants appear as specializations of invariants of quiver moduli spaces, on the other hand knot invariants (HOMFLY-PT polynomials, or LMOV invariants) can be defined in more general families, labeled by arbitrary Young diagrams (not just symmetric Young diagrams, which appear in (\ref{PxC})). It is desirable to understand how such more general invariants are related to, or could be extracted from, the data of quiver moduli spaces. 

We also note that, as argued in \cite{Kontsevich:2010px}, the Cohomological Hall Algebra associated to a quiver should be identified with the algebra of BPS states \cite{Harvey:1996gc}. Furthermore, the generating functions (\ref{PxC}) take the form of products of $q$-series that appear in Nahm conjectures \cite{Nahm}, which suggests their relation to conformal field theories and integrability. All these issues are worth thorough further investigation.


\subsection{The strategy and $q$-identities}

In order to determine a quiver corresponding to a given knot, we have to rewrite the generating function of colored HOMFLY-PT polynomials of this knot in the form (\ref{PxC}). Colored HOMFLY-PT polynomials, which are known for various knots, can be written in terms of sums involving $q$-Pochhammer and $q$-binomial symbols \cite{superA,FGSS}, as e.g. in the expression (\ref{eq:Torus knots from FGS12}). However, in general in such expressions the number of summations is smaller than the number of terms in the superpolynomial, and it is not obvious that such sums can be rewritten in the form (\ref{PxC}), which involves as many summations as the number of terms in the superpolynomial. Therefore some algebraic manipulations are necessary in order to rewrite such formulas in the form that involves an appropriate number of additional summations, and in addition includes appropriate $q$-Pochhammer symbols in the denominator. To achieve this we take advantage of the following lemmas.

\begin{lemma}\label{lem1}
For any $d_1,\ldots,d_k\ge 0$, we have:
\begin{align}
\frac{(x;q)_{d_1+\ldots+d_k}}{(q;q)_{d_1}\cdots (q;q)_{d_k}}
=& \sum\limits_{\alpha_1+\beta_1=d_1}\sum\limits_{\alpha_2+\beta_2=d_2}\cdots\sum\limits_{\alpha_k+\beta_k=d_k}  
\frac{1}{(q;q)_{\alpha_1}\cdots(q;q)_{\alpha_k}(q;q)_{\beta_1}\cdots(q;q)_{\beta_k}}  \times  \nonumber \\
& \times (-x)^{{\alpha_1+\ldots+\alpha_k}} q^{\frac{1}{2}(\alpha_1^2+\ldots+\alpha_k^2)}q^{\sum_{i=1}^{k-1} \alpha_{i+1} (d_1+\ldots+d_i)}q^{-\frac{1}{2}(\alpha_1+\ldots+\alpha_k)}.    \label{lem1-eq}
\end{align}
\end{lemma}

\textbf{Proof:}\\

First, note that
\be
(x;q)_{d_1+\ldots+d_k}=(x;q)_{d_1}(xq^{d_1};q)_{d_2}(xq^{d_1+d_2};q)_{d_3}\ldots(xq^{d_1+\ldots+d_{k-1}};q)_{d_k}.
\ee
By expanding each of $k$ $q$-Pochhammers on the right hand side by using the quantum binomial identity
\be
(x;q)_n= \sum\limits_{\alpha=0}^n (-x)^{\alpha} q^{\frac{1}{2}\alpha(\alpha-1)} \frac{(q;q)_{n}}{(q;q)_{\alpha} (q;q)_{n-\alpha}} 
= \sum\limits_{\alpha+\beta=n} (-x)^{\alpha} q^{\frac{1}{2}\alpha(\alpha-1)}\frac{(q;q)_{n}}{(q;q)_{\alpha}(q;q)_{\beta}},  \label{qpoch-sum}
\ee
we obtain
\begin{align}(x;q)_{d_1+\ldots+d_k}=  
&\sum\limits_{\alpha_1+\beta_1=d_1}
\sum\limits_{\alpha_2+\beta_2=d_2}
\cdots
\sum\limits_{\alpha_k+\beta_k=d_k}  
\frac{(q;q)_{d_1}}{(q;q)_{\alpha_1}(q;q)_{\beta_1}}\frac{(q;q)_{d_2}}{(q;q)_{\alpha_2}(q;q)_{\beta_2}}\cdots\frac{(q;q)_{d_k}}{(q;q)_{\alpha_k}(q;q)_{\beta_k}}
\nonumber \\
& \times (-x)^{\alpha_1+\ldots+\alpha_k} 
q^{{\frac{1}{2}}(\alpha_1^2+\ldots+\alpha_k^2-\alpha_1-\ldots-\alpha_k)}
q^{\sum_{i=1}^{k-1}{ \alpha_{i+1} (d_1+\ldots+d_i)}}
\end{align}
which proves the lemma.   \kraj

Lemma \ref{lem1} enables rewriting the expression of the form 
\be
\frac{(x;q)_{d_1+\ldots+d_k}}{(q;q)_{d_1}\cdots (q;q)_{d_k}}
\ee
as a sum of terms
\be
\frac{1}{(q;q)_{\alpha_1}\cdots(q;q)_{\alpha_k}(q;q)_{\beta_1}\cdots(q;q)_{\beta_k}}
\ee
weighted simply by linear and quadratic powers of $q$.  
In this way (at least in some cases) we can introduce additional summations in expressions for colored HOMFLY-PT polynomials, in order to bring them into the form of the quiver generating series (\ref{PxC}). In more complicated situations, generalizing the relation
\begin{equation}\label{pomoc1}
(x;q)_{a+b}=(x;q)_a(xq^a;q)_b,
\end{equation}
we can take advantage of the following lemma which enables rewriting certain $q$-binomial coefficients. Recall that throughout the paper we are using the convention ${n\brack k}=\frac{(q^2;q^2)_n}{(q^2;q^2)_k(q^2;q^2)_{n-k}}$, cf. (\ref{q-binomial}).
\begin{lemma}\label{lem2}
For nonnegative integers $a,b$ and $k$ we have
\begin{equation}\label{prva}
{ a+b \brack k}=\sum_{i+j=k} q^{2(a-i)(k-i)} {a\brack i} {b\brack j}.
\end{equation}
More generally, let $a_1,\ldots,a_m$, $m\ge 1$, and $k_1,\ldots,k_p$, $p\ge 1$, be nonnegative integers. Then
\begin{align}
{a_1+a_2\brack k_1} {k_1\brack k_2} \cdots {k_{p-1}\brack k_p} =& \sum_{i_1+j_1=k_1}\sum_{i_2+j_2=k_2}\cdots
\sum_{i_p+j_p=k_p}    {a_1\brack i_1} {i_1\brack i_2}\cdots { i_{p-1}\brack i_p} \,  {a_2\brack j_1 }{ j_1\brack j_2 } \cdots { j_{p-1}\brack j_p}  \nonumber \\
& \times q^{2\left((a_1-i_1)(k_1-i_1)+(i_1-i_2)(k_2-i_2)+\ldots+(i_{p-1}-i_p)(k_p-i_p)\right)}  ,
\end{align}
\begin{align}
{ a_1+a_2+\ldots+a_m\brack k}=& \sum_{i_1+i_2+\ldots+i_m=k}  { a_1\brack i_1 }{ a_2\brack i_2 }\cdots { a_{m}\brack i_m } \nonumber \\
& \times q^{2\left((a_1-i_1)(k-i_1)+(a_2-i_2)(k-i_1-i_2)+\ldots+(a_m-i_m)(k-i_1-i_2-\ldots-i_m)\right)}.
\end{align}
Furthermore
\begin{align}
&{ a_1+a_2+\ldots+a_m\brack k_1 } {k_1\brack k_2} \cdots { k_{p-1}\brack k_p }= \nonumber \\
& =  \sum_{i^1_1+\ldots+i^1_m=k_1} \sum_{i^2_1+\ldots+i^2_m=k_2}\cdots \sum_{i^p_1+\ldots+i^p_m=k_p}
q^{X(\underline{a},\underline{i^1},k_1)+X(\underline{i^1},\underline{i^2},k_2)+\ldots+X(\underline{i^{p-1}},\underline{i^p},k_p)} \\
&\qquad\quad\times  { a_1\brack i^1_1 }{  i^1_1\brack i^1_2 } \cdots { i^1_{p-1}\brack i^1_p } \ { a_2\brack i^2_1 }{ i^2_1\brack i^2_2} \cdots { i^2_{p-1}\brack i^2_p }\cdots {a_m\brack i^m_1 }{ i^m_1\brack i^m_2 }\cdots
{ i^m_{p-1}\brack i^m_p } \nonumber
\end{align}
where
$$
X(\underline{a},\underline{i},k)={2\left((a_1-i_1)(k-i_1)+(a_2-i_2)(k-i_1-i_2)+\ldots+(a_m-i_m)(k-i_1-i_2-\ldots-i_m)\right)},
$$
for sequences $\underline{a}=(a_1,\ldots,a_m)$ and $\underline{i}=(i_1,\ldots,i_m)$. 
\end{lemma}
\textbf{Proof:}
These relations follow after straightforward calculations. For example, from the $q$-Pochhammer relation (\ref{pomoc1}) we get
\be
\sum_{k=0}^{a+b}(-x)^k q^{k^2-k} { a+b\brack k } = \sum_{i=0}^a (-x)^i q^{i^2-i} { a\brack i } \sum_{j=0}^b (-x)^j q^{2aj} q^{j^2-j} { b\brack j }.
\ee
By matching the powers of $x$ on both sides of this equation we get the relation (\ref{prva}). The remaining three equalities can now be obtained by induction, using (\ref{prva}). \kraj

One can use the above lemma for example after rewriting the following product of binomial coefficients in terms of the $q$-Pochhammer symbols 
\begin{equation}
{ a\brack k_1}{ k_1\brack k_2} \cdots { k_{p-1}\brack k_p}=\frac{(q^2;q^2)_a}{(q^2;q^2)_{a-k_1}(q^2;q^2)_{k_1-k_2}\cdots(q^2;q^2)_{k_{p-1}-k_p}(q^2;q^2)_{k_p}}.
\end{equation}
The right hand side is the quotient of the $q$-Pochhammer of length $a$ by the product of $p$ $q$-Pochhammers, whose lengths sum up also to $a$. As we explained above, this can be further transformed into a sum involving $q$-Pochhammers only in the denominator, weighted only by linear and quadratic powers of $q$, which is then of the required quiver form (\ref{PxC}).


\subsection{Reduced vs. unreduced invariants}    \label{ssec-unreduced}

The values of parameters $l_i, a_i$ and $C_{i,i}$ in (\ref{PxC}) depend on the choice of normalization of $\overline{P}_r(a,q)$. The values mentioned in Conjecture \ref{conj2} arise when the normalization includes only the denominator $(q^2;q^2)_r$ of the colored HOMFLY-PT polynomial of the unknot, i.e. 
\be
\overline{P}_r(a,q)=\frac{P_r(a,q) }{ (q^2;q^2)_r}.   \label{barPr-qq}
\ee
In this case the values of $l_i, a_i$ and $C_{i,i}$ are related to reduced and uncolored HOMFLY-PT homology and superpolynomial. On the other hand, a more familiar normalization that involves the full unknot polynomial  (\ref{Pbar-unknot})
\be
\overline{P}_r(a,q)=a^{-r}q^r \frac{(a^2;q^2)_r}{(q^2;q^2)_r} P_r(a,q)  \label{barPr-unknot}
\ee
leads to a twice larger quiver, which encodes information about unreduced HOMFLY-PT homology, whose Poincar{\'e} polynomial is obtained by multiplying the (reduced) superpolynomial by $a^{-1}q(1+a^2 t)$. Suppose that colored polynomials normalized as in (\ref{barPr-qq}) lead to a quiver encoded in a matrix $C$. Multiplying (\ref{barPr-qq}) by an additional factor $a^{-r}q^r (a^2;q^2)_r$, we can use (\ref{lem1-eq}) to deal with the additional $q$-Pochhammer $(a^2;q^2)_r$. Introducing new summation variables $\alpha_i$ and $\beta_i$, such that $d_i = \alpha_i + \beta_i$, the expression (\ref{PxC}) is replaced by another summation which is also of the required form (\ref{PxC}), which however involves summations over $\alpha_i$ and $\beta_i$ with summands involving the following factors of $q$ in quadratic powers of summation variables
\be
q^{\sum_{i,j} C_{i,j}(\alpha_i+\beta_i)(\alpha_j+\beta_j)} q^{\alpha_1^2 +\ldots \alpha_m^2} q^{2\sum_{i=1}^{m-1} \alpha_{i+1}(d_1 + \ldots d_i) } .
\ee
The exponent of $q$ in this expression can be rewritten as
\be
\sum_{i,j} C_{i,j} \beta_i \beta_j + \sum_{i,j} (C_{i,j}+1) \alpha_i \alpha_j + 2\sum_{i\leq j} C_{i,j} \alpha_i \beta_j + 2 \sum_{i>j} (C_{i,j}+1) \alpha_i \beta_j.  \label{C-twice}
\ee
This expression also encodes a quiver, which is however twice larger than $C$, and which decomposes into two parts: one which looks like the original quiver encoded in $C$ (determined by the first term $\sum_{i,j} C_{i,j} \beta_i \beta_j$), and another one which looks like the original quiver framed by 1 (as determined by the second term $\sum_{i,j} (C_{i,j}+1) \alpha_i \alpha_j$). These two subquivers are connected by arrows, whose structure is given by the last two summations in (\ref{C-twice}).


\section{Case studies}   \label{sec-results}

In this section we illustrate our claims and conjectures in various examples. We show how to rewrite generating functions of known colored HOMFLY-PT polynomials in the form (\ref{PxC}) and identify corresponding quivers. In particular this automatically proves the LMOV conjecture (for symmetric representations) for the knots under consideration. Furthermore, assuming that generating functions of colored HOMFLY-PT polynomials should be of the form (\ref{PxC}), we derive previously unknown formulas for such polynomials for $6_2$ and $6_3$ knots, as well as for $(3,7)$ torus knot.


\subsection{Unknot}

\begin{figure}[b]
\begin{center}
\includegraphics[width=0.35\textwidth]{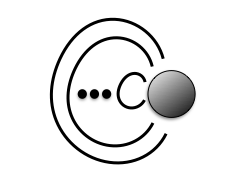} 
\caption{Quiver with one vertex and $f$ loops, encoding extremal framed unknot invariants (equivalently open topological string amplitudes for branes in $\mathbb{C}^3$ geometry).}  \label{fig-m-loop}
\end{center}
\end{figure}

The (unreduced) colored HOMFLY-PT polynomial for the unknot takes the form
\be
\overline{P}_r(a,q) = a^{-r}q^r \frac{(a^2;q^2)_r}{(q^2;q^2)_r}.    \label{Pbar-unknot}
\ee
First, consider just the denominator of this expression, which includes a single $q$-Pochhammer. This is equivalent to the simpler (reduced) normalization discussed in section \ref{ssec-unreduced}, and up to the $q^r$ factor it agrees with the extremal (bottom row) unknot HOMFLY-PT polynomial. More generally, the generating series of the reduced colored HOMFLY-PT polynomials of the $f$-framed unknot takes the form
\be
P(x) = \sum_{r=0}^\infty x^r \frac{q^{f(r^2-r)}}{(q^2;q^2)_r},    \label{Px-unknot}
\ee
which essentially agrees with the motivic generating series associated to a quiver consisting of one vertex and $f$ loops, shown in figure \ref{fig-m-loop}. These are prototype and important examples of quivers, and properties of their moduli spaces were discussed in \cite{COM:8276935,Rei12}. The relation of this family of quivers to LMOV invariants of framed unknot (equivalently extremal invariants of twist knots, or open topological string amplitudes for branes in $\mathbb{C}^3$ geometry) was presented in \cite{Kucharski:2016rlb}, and discussed also in \cite{Luo:2016oza,Zhu:2017lsn}. 

Consider now the generating function of the full unknot invariants (\ref{Pbar-unknot}) -- or equivalently open topological string amplitudes for branes in the resolved conifold geometry. Using (\ref{qpoch-sum}), this generating function can be rewritten as
\begin{align}
P(x) &= \sum_{r=0}^{\infty} x^r a^{-r}q^r \frac{(a^2;q^2)_r}{(q^2;q^2)_r} = \sum_{d_1,d_2=0}^{\infty} x^{d_1+d_2} \frac{(-1)^{d_1} a^{d_1-d_2} q^{d_1^2+d_2} }{(q^2;q^2)_{d_1} (q^2;q^2)_{d_2}} = \nonumber \\
& = \Big(  \sum_{d_1=0}^{\infty} x^{d_1}  \frac{(-1)^{d_1} a^{d_1} q^{d_1^2} }{(q^2;q^2)_{d_1}} \Big) 
\Big(  \sum_{d_2=0}^{\infty} x^{d_2}  \frac{a^{-d_2} q^{d_2} }{(q^2;q^2)_{d_2}} \Big) = \frac{(xaq;q^2)_{\infty}}{(xq/a;q^2)_{\infty}}.   \label{Px-unknot-a}
\end{align}
From the expression in the first line, or simply taking advantage of (\ref{C-twice}), we find that the corresponding quiver can be interpreted as a twice larger quiver associated to (\ref{Px-unknot}); this larger quiver consists of two disconnected vertices, with a single loop associated to one vertex (labeled by $d_1$). The final factorization into the ratio of two quantum dilogarithms means that there are only two non-zero LMOV (or motivic Donaldson-Thomas) invariants, which is a well known statement for the unknot \cite{OoguriV,Kucharski:2016rlb}. 

More generally, including the framing dependence (\ref{framing}) in (\ref{Px-unknot-a}) results in a quiver with additional loops and arrows, as in (\ref{C-frame}). Contrary to the unframed case (\ref{Px-unknot-a}), such expressions do not factorize into a finite number of quantum dilogarithms, and they would encode an infinite number of LMOV invariants.


\subsection{Trefoil and cinquefoil knots}   \label{ssec-3151}

We now illustrate how to identify a quiver corresponding to a knot in the example of the trefoil knot, i.e. the $(2,3)$  torus knot (also denoted $T_{2,3}$ or $3_1$), whose reduced colored HOMFLY-PT polynomials arise by setting $t=-1$ in (\ref{Pr-31}) 
\be
P_r(a,q)=  \frac{a^{2r}}{q^{2r}}\sum_{k=0}^r {r \brack k} q^{2k(r+1)} \prod_{i=1}^k  (1-a^2q^{2(i-2)}),    \label{Pr-31-homfly}
\ee
with the $q$-binomial ${r\brack k}$ given in (\ref{q-binomial}). Using (\ref{qpoch-sum}), the $q$-binomial together with the last product in (\ref{Pr-31-homfly}) take the form
$$
{r \brack k} \big(\frac{a^2}{q^2};q^2\big)_k = \sum_{i=0}^k \frac{(q^2;q^2)_r  \big( - \frac{a^2}{q^2} \big)^i q^{i(i-1)} }{(q^2;q^2)_{r-k} (q^2;q^2)_i (q^2;q^2)_{k-i}}.
$$
Introducing 
\be
r=d_1+d_2+d_3, \quad k=d_2+d_3, \quad  i=d_3,
\ee
with $d_i\geq 0$, and normalizing $P_r(a,q)$ by $(q^2;q^2)_r$, the generating function (\ref{Pz2}) takes the form
\be 
P(x) = \sum_{r=0}^{\infty} \frac{P_r(a,q)}{(q^2;q^2)_r} x^r = \sum_{d_1,d_2,d_3\geq 0} \frac{q^{\sum_{i,j} C^{T_{2,3}}_{i,j} d_i d_j  -2d_1-3d_3} (-1)^{d_3} a^{2d_1+2d_2+4d_3}   }{(q^2;q^2)_{d_1} (q^2;q^2)_{d_2} (q^2;q^2)_{d_3}}   x^{d_1+d_2+d_3},  \label{PxC-31}
\ee
where 
\arraycolsep 3pt
\be
C^{T_{2,3}} = \left[
\begin{array}{ccc}
0 & 1 & 1 \\
1 & 2 & 2 \\
1 & 2 & 3 \\
\end{array}
\right]    \label{trefoil-quiver}
\ee
The expression (\ref{PxC-31}) is indeed of the form (\ref{PxC}), with the corresponding quiver shown in figure \ref{fig-trefoil}. Vertices of this quiver, as stated in the previous section, correspond to generators of HOMFLY-PT homology. The diagonal elements $(0,2,3)$ of the matrix $C$ (representing numbers of loops at vertices of the quiver) indeed agree with homological degrees encoded in the uncolored superpolynomial (\ref{P1-31}), coefficients $l_i=-2,0,3$ of linear terms in $d_i$ in the power of $q$ in (\ref{PxC-31}) are given by $l_i=q_i-t_i$, coefficients $a_i=2,2,4$ in the power of $a$ agree with $a$-degrees of generators of HOMFLY-PT homology, and the additional minus sign $(-1)^{d_3}$ is determined by just one generator with odd $t$-degree $t_3=3$ (which is manifest in figure \ref{diagrams}).

Let us also discuss the normalization of the colored HOMFLY-PT polynomials by the full unknot invariant, following section \ref{ssec-unreduced}. Multiplying each summand in (\ref{PxC-31}) proportional to $x^r$ by $a^{-r}q^r (a^2;q^2)_r$ and taking advantage of (\ref{lem1-eq}), we get the generating function
\begin{align}
P(x) = & \sum_{r=0}^{\infty} x^r   \sum\limits_{\alpha_1+\alpha_2+\alpha_3+\beta_1+\beta_2+\beta_3=r}
\frac{  q^{\alpha_1^2+\alpha_2^2+\alpha_3^2}q^{\alpha_2 d_1+\alpha_{3} (d_1+d_2)}q^{-\alpha_1-\alpha_2-\alpha_3}     }{(q^2;q^2)_{\alpha_1}
(q^2;q^2)_{\alpha_2}(q^2;q^2)_{\alpha_3}(q^2;q^2)_{\beta_1}(q^2;q^2)_{\beta_2}(q^2;q^2)_{\beta_3}} \nonumber \\
&\quad  \times  (-1)^{d_3}a^{d_1+d_2+3d_3} q^{-d_1+d_2-2d_3}q^{2d_2^2+3d_3^2+2(d_1d_2+d_1d_3+2d_2d_3)} (-a^2 )^{\alpha_1+\alpha_2+\alpha_3},  \label{PxC-31-bis}
\end{align}
where $d_i=\alpha_i+\beta_i$, $i=1,2,3$. The form of the corresponding quiver can be read off from powers of $q$ in this generating function, or simply from the transformation (\ref{C-twice}) applied to the quiver (\ref{trefoil-quiver}). Ultimately we find a quiver with 6 nodes, whose structure, in the basis ordered as $(\beta_1,\alpha_1,\beta_2,\alpha_2,\beta_3,\alpha_3)$, is encoded in the matrix of the form
\be
C^{T_{2,3}}_{unreduced} =
\left[
\begin{array}{cccccc}
0\ & 0\ & 1\ & 2\ & 1\ & 2 \\
0\ & 1\ & 1\ & 2\ & 1\ & 2 \\
1\ & 1\ & 2\ & 2\ & 2\ & 3 \\
2\ & 2\ & 2\ & 3\ & 2\ & 3 \\
1\ & 1\ & 2\ & 2\ & 3\ & 3 \\
2\ & 2\ & 3\ & 3\ & 3\ & 4 \\
\end{array}  \label{C-trefoil-bis}
\right]
\ee
As expected, the information about unreduced HOMFLY-PT homology is encoded in this quiver and the expression (\ref{PxC-31-bis}). For the trefoil this homology has 6 generators (obtained by multiplying the reduced superpolynomial by $a^{-1}q(1+a^2 t)$), with $t$-degrees $0, 1, 2, 3, 3, 4$, which indeed appear as diagonal elements in (\ref{C-trefoil-bis}). More generally, $(q,t)$-degrees of these six generators are $(-1,0),(-1,1),(3,2),(3,3),(1,3),(1,4)$, and the differences $q_i-t_i$ in (\ref{linear}) also match the coefficients of the linear term in the power of $q$ in (\ref{PxC-31-bis}), which are of the form $-\beta_1-2\alpha_1+\beta_2-2\beta_3-3\alpha_3$.

Let us consider a more involved example of the cinquefoil  $(2,5)$ torus knot (also denoted $T_{2,5}$ or $5_1$). Its colored HOMFLY-PT polynomials are obtained as the $p=2$ case of (\ref{eq:Torus knots from FGS12}) and their generating function, normalized by $(q^2;q^2)_r$, takes the form
\be
P(x)=  \sum_{r=0}^{\infty} \frac{x^{r}a^{4r}q^{-4r}}{(q^2;q^2)_r}  \sum_{0\leq k_{2}\leq k_{1}\leq r}
{r \brack k_1} {k_1 \brack k_2}   q^{2(2r+1)(k_{1}+k_{2})-2rk_{1}-2k_{1}k_{2}}(a^{2}q^{-2};q^{2})_{k_{1}}. \label{eq:5_1 general formula}
\ee
Rewriting the last $q$-Pochhammer symbol in this expression using (\ref{qpoch-sum}), and then taking advantage of (\ref{prva}), we get
\begin{align}
P(x)= & \sum_{r=0}^{\infty} \frac{x^{r} }{(q^2;q^2)_r} \sum_{0\leq k_{2}\leq k_{1}\leq r}\sum_{0\leq\alpha_{1}\leq k_{1}}
{r \brack k_1}{k_1 \brack k_2} {k_1 \brack \alpha_1} 
\times\nonumber \\
 & \times (-1)^{\alpha_1}a^{2\alpha_1+4r} q^{2(2r+1)(k_{1}+k_{2})-2rk_{1}-2k_{1}k_{2} + \alpha_{1}^{2}-3\alpha_{1} -4r} = \nonumber \\
= & \sum_{r=0}^{\infty} \frac{x^{r} }{(q^2;q^2)_r}  \sum_{0\leq k_{2}\leq k_{1}\leq r}\sum_{0\leq\alpha_{1}\leq k_{1}}\sum_{0\leq\alpha_{2}\leq k_{2}}{r \brack k_1} {k_1 \brack \alpha_1} {k_{1}-\alpha_{1} \brack  k_{2}-\alpha_{2}} {\alpha_{1} \brack \alpha_{2}}
\times\nonumber \\
 & \times(-1)^{\alpha_{1}} a^{4r+2\alpha_{1}}q^{-4r+2(2r+1)(k_{1}+k_{2})-2rk_{1}-2k_{1}k_{2}+\alpha_{1}^{2}-3\alpha_{1}+2(\alpha_{1}-\alpha_{2})(k_{2}-\alpha_{2})}\label{eq:5_1 generating function}
\end{align}
(with the condition $\alpha_{2}\leq\alpha_{1}, k_{2}-\alpha_{2}\leq k_{1}-\alpha_{1}$). After the change of variables
\be
d_{1} =r-k_{1}, \quad d_{2} =k_{1}-\alpha_{1}-(k_{2}-\alpha_{2}), \quad d_{3} =\alpha_{1}-\alpha_{2},
\quad d_{4} =k_{2}-\alpha_{2},\quad d_{5} =\alpha_{2},
\ee
we finally get
\begin{align}
P(x)= & \sum_{d_{1},d_{2},\ldots,d_{5}\geq0}q^{\sum_{i,j}C_{i,j}^{T^{2,5}}d_{i}d_{j}}  \frac{x^{d_{1}+d_{2}+\cdots+d_{5}}}{(q^{2};q^{2})_{d_{1}}(q^{2};q^{2})_{d_{2}}\cdots(q^{2};q^{2})_{d_{5}}}\times\label{eq:5_1 quiver version}\\
 & \times(-1)^{d_{3}+d_{5}}a^{4d_{1}+4d_{2}+6d_{3}+4d_{4}+6d_{5}}q^{-4d_{1}-2d_{2}-5d_{3}-3d_{5}},\nonumber 
\end{align}
where the matrix
\begin{equation}
C^{T_{2,5}}=\left[\begin{array}{ccccc}
0\ & 1\ & 1\ & 3\ & 3 \\
1\ & 2\ & 2\ & 3\ & 3 \\
1\ & 2\ & 3\ & 4\ & 4 \\
3\ & 3\ & 4\ & 4\ & 4 \\
3\ & 3\ & 4\ & 4\ & 5
\end{array}\right]    \label{51-quiver}
\end{equation}
represents the quiver corresponding to the $(2,5)$ torus knot.



\subsection{$(2,2p+1)$  torus knots}        \label{ssec-torus22p1}

Colored HOMFLY-PT polynomials for $(2,2p+1)$ torus knots (also denoted $T_{2,2p+1}$) can be obtained as the $t=-1$ specialization of the following colored superpolynomials, determined in \cite{FGS,FGSS}
\begin{align}
P_{S^{r}}^{T_{2,2p+1}}(a,q,t)= & a^{2pr}q^{-2pr}\sum_{0\leq k_{p}\leq\ldots\leq k_{2}\leq k_{1}\leq r} {r \brack k_1} {k_1 \brack k_2} \cdots { k_{p-1}\brack  k_{p}} \times\nonumber \\
 & \times q^{2\sum_{i=1}^{p}((2r+1)k_{i}-k_{i-1}k_{i})}t^{2(k_{1}+k_{2}+\ldots+k_{p})}\prod_{i=1}^{k_{1}}\big(1+a^{2}q^{2(i-2)}t\big).    \label{eq:Torus knots from FGS12}
\end{align}
These expressions can be transformed to the form (\ref{PxC}) recursively, generalizing the step between trefoil and cinquefoil knots presented in the previous section. In order to determine the form of the generating function (\ref{PxC}) and the quiver for arbitrary $(2,2p+1)$ torus knot, we analyze first the following three modifications in the expression for colored HOMFLY-PT polynomials, when $p$ is changed to $p+1$ 
\begin{align}
a^{2pr}q^{-2pr}\ &\mapsto\  a^{2(p+1)r}q^{-2(p+1)r}   \label{mod1}   \\
\sum_{0\leq k_{p}\leq\ldots\leq k_{1}\leq r} {r\brack   k_{1}}
\cdots {k_{p-1}\brack k_{p}}\ &\mapsto\ \sum_{0\leq k_{p+1}\leq\ldots\leq k_{1}\leq r}{ r\brack k_{1}}
\cdots {k_{p}\brack  k_{p+1}} \label{mod2}   \\
q^{2\sum_{i=1}^{p}\left[(2r+1)k_{i}-k_{i-1}k_{i}\right]}\ &\mapsto\ q^{2\sum_{i=1}^{p}\left[(2r+1)k_{i}-k_{i-1}k_{i}\right]+2(2r+1)k_{p+1}-2k_{p}k_{p+1}}   \label{mod3}
\end{align}
These transformations generalize the relation between trefoil and cinquefoil knots, which we discussed in section \ref{ssec-3151}, and which corresponds to changing $p=1$ to $p=2$.

The first modification (\ref{mod1}) only affects the change of variables leading to the generating function of the quiver, but not the form of the quiver. 

In the second transformation (\ref{mod2}) a new variable $k_{p+1}$ and an additional $q$-binomial ${k_{p}\brack k_{p+1}}$ are introduced. Let us discuss first the special $p=1$ case of trefoil and cinquefoil knots. As already analyzed above, in this case, in the generating series for the cinquefoil knot, we split ${k_{p}\brack k_{p+1}}\equiv {k_{1}\brack k_{2}}$ into ${k_{p}-\alpha_{p}\brack  k_{p+1}-\alpha_{p+1}} {\alpha_{p}\brack \alpha_{p+1}}$ and changed variables accordingly
\begin{align}
d_{1} & =r-k_{1} &  &  & d_{1} & =r-k_{1}\nonumber \\
d_{2} & =k_{1}-\alpha_{1} & \longmapsto &  & d_{2} & =k_{1}-\alpha_{1}-(k_{2}-\alpha_{2})\nonumber \\
d_{3} & =\alpha_{1} &  &  & d_{3} & =\alpha_{1}-\alpha_{2}\\
 &  &  &  & d_{4} & =k_{2}-\alpha_{2}\nonumber \\
 &  &  &  & d_{5} & =\alpha_{2}\nonumber 
\end{align}
This is equivalent to the following modification of summation variables in the quiver generating series
\begin{align}
r & =d_{1}+d_{2}+d_{3} &  &  & r & =d_{1}+(d_{2}+d_{4})+(d_{3}+d_{5})\nonumber \\
k_{1} & =d_{2}+d_{3} & \longmapsto &  & k_{1} & =(d_{2}+d_{4})+(d_{3}+d_{5})\label{eq:change of vars 5_1}\\
\alpha_{1} & =d_{3} &  &  & \alpha_{1} & =(d_{3}+d_{5})\nonumber 
\end{align}
which means that the matrix representing the cinquefoil quiver is obtained from the one for the trefoil quiver by copying the first and the second column and row, respectively, into the third and the fourth column and row. In addition, changing ${k_{1}\brack k_{2}}$ into ${k_{p}-\alpha_{p}\brack  k_{p+1}-\alpha_{p+1}} {\alpha_{p}\brack \alpha_{p+1}}$ introduces a new term in $\sum_{i,j}C_{i,j}d_{i}d_{j}$ of the form $2(\alpha_{1}-\alpha_{2})(k_{2}-\alpha_{2})=2d_{3}d_{4}$, which means that the matrix elements $C_{3,4}$ and $C_{4,3}$ are increased by $1$.

Generalizing the above transformation and splitting ${k_{p}\brack k_{p+1}}$ into ${k_{p}-\alpha_{p}\brack  k_{p+1}-\alpha_{p+1}} {\alpha_{p}\brack \alpha_{p+1}}$ for arbitrary $p$, the relation (\ref{eq:change of vars 5_1}) is replaced by
\begin{align}
r & =d_{1}+d_{2}\ldots+d_{2p}+d_{2p+1} &  &  & r & =d_{1}+d_{2}\ldots+(d_{2p}+d_{2p+2})+(d_{2p+1}+d_{2p+3})\nonumber \\
k_{1} & =d_{2}+\ldots+d_{2p}+d_{2p+1} &  &  & k_{1} & =d_{2}+\ldots+(d_{2p}+d_{2p+2})+(d_{2p+1}+d_{2p+3})\nonumber \\
k_{2} & =d_{4}+\ldots+d_{2p}+d_{2p+1} &  &  & k_{2} & =d_{4}+\ldots+(d_{2p}+d_{2p+2})+(d_{2p+1}+d_{2p+3})\nonumber \\
 & \vdots &  &  &  & \vdots\nonumber \\
k_{p} & =d_{2p}+d_{2p+1} & \longmapsto &  & k_{p} & =(d_{2p}+d_{2p+2})+(d_{2p+1}+d_{2p+3})\nonumber \\
\alpha_{1} & =d_{3}+d_{5}+\ldots+d_{2p+1} &  &  & \alpha_{1} & =d_{3}+d_{5}+\ldots+(d_{2p+1}+d_{2p+3})\nonumber \\
\alpha_{2} & =d_{5}+\ldots+d_{2p+1} &  &  & \alpha_{2} & =d_{5}+\ldots+(d_{2p+1}+d_{2p+3})\nonumber \\
 & \vdots &  &  &  & \vdots\label{eq:change of vars general}\\
\alpha_{p} & =d_{2p+1} &  &  & \alpha_{p} & =(d_{2p+1}+d_{2p+3})\nonumber 
\end{align}
so that columns and rows of number $2p$ and $2p+1$ are copied respectively to those of number $2p+2$ and $2p+3$, and matrix elements $C_{2p+1,2p+2}$ and $C_{2p+2,2p+1}$ are increased by $1$.

Finally, the third transformation (\ref{mod3}) modifies the change of variables and adds $4rk_{p+1}-2k_{p}k_{p+1}$
to the sum $\sum{}_{i,j}C_{i,j}d_{i}d_{j}$. In the special case of $p=1$ we have
\begin{align}
4rk_{2}-2k_{1}k_{2}&=  4(r-k_{1})k_{2}+2(k_{1}-k_{2})k_{2}+2k_{2}^{2} = \nonumber \\
&=  4d_{1}\left(d_{4}+d_{5}\right)+2\left(d_{2}+d_{3}\right)\left(d_{4}+d_{5}\right)+2\left(d_{4}+d_{5}\right)^{2},
\end{align}
which means that $C_{1,4}, C_{1,5}, C_{4,5}$ (and transposed matrix elements)
and $C_{4,4}$, $C_{5,5}$ increase by $2$,  and $C_{2,4}, C_{2,5}, C_{3,4}$ and $C_{3,5}$ (and transposed elements) increase by $1$. For general $p$
\begin{align}
4rk_{p+1}-2k_{p}k_{p+1} &=  4(r-k_{p})k_{p+1}+2(k_{p}-k_{p+1})k_{p+1}+2k_{p+1}^{2} =\nonumber \\
&=  4\left(d_{1}+\ldots+d_{2p-1}\right)\left(d_{2p+2}+d_{2p+3}\right) + \\
&\quad   +2\left(d_{2p}+d_{2p+1}\right)\left(d_{2p+2}+d_{2p+3}\right)+2\left(d_{2p+2}+d_{2p+3}\right)^{2},\nonumber 
\end{align}
which means increasing
$C_{1,2p+2}, \ldots, C_{2p-1,2p+2}$, $C_{1,2p+3}, \ldots,
C_{2p-1,2p+3}$, $C_{2p+2,2p+3}$ (and transposed elements) and $C_{2p+2,2p+2}$,
$C_{2p+3,2p+3}$ by $2$, as well as increasing $C_{2p,2p+2}$, $C_{2p+1,2p+2}$,
$C_{2p,2p+3}$, $C_{2p+1,2p+3}$ (and transposed elements) by $1$.

To sum up, once we know a matrix $C^{T_{2,2p+1}}$ representing a quiver for the $(2,2p+1)$ torus knot, the matrix $C^{T_{2,2p+3}}$ for a quiver associated to the $(2,2p+3)$ torus knot is obtained by copying columns and rows of the number $2p$ and $2p+1$ to $2p+2$ and $2p+3$ respectively, and increasing elements in the last two columns (and rows) by $2$, except for $C_{2p,2p+2}$, $C_{2p,2p+3}$, and $C_{2p+1,2p+3}$ (and transposed elements) that are increased by $1$. The solution of this recursion, for an arbitrary $(2,2p+1)$ torus knot, takes the form
\begin{equation}
C^{T_{2,2p+1}}=\left[\begin{array}{ccccccc}
F_{0} & F_{1} & F_{2} & F_{3} & \cdots & F_{p-1} & F_{p}\\
F_{1}^{T} & D_{1} & U_{2} & U_{3} & \cdots & U_{p-1} & U_{p}\\
F_{2}^{T} & U_{2}^{T} & D_{2} & U_{3} & \cdots & U_{p-1} & U_{p}\\
F_{3}^{T} & U_{3}^{T} & U_{3}^{T} & D_{3} & \cdots & U_{p-1} & U_{p}\\
\vdots & \vdots & \vdots & \vdots & \ddots & \vdots & \vdots\\
F_{p-1}^{T} & U_{p-1}^{T} & U_{p-1}^{T} & U_{p-1}^{T} & \cdots & D_{p-1} & U_{p}\\
F_{p}^{T} & U_{p}^{T} & U_{p}^{T} & U_{p}^{T} & \cdots & U_{p}^{T} & D_{p}
\end{array}\right]    \label{C-torus22p1}
\end{equation}
with the following block entries
\be
F_{0}=\left[0\right],\qquad
F_{k}=\left[\begin{array}{cc}
2k-1 & 2k-1\end{array}\right],
\qquad
D_{k}=\left[\begin{array}{cc}
2k & 2k\\
2k & 2k+1
\end{array}\right], \qquad
U_{k}=\left[\begin{array}{cc}
2k-1 & 2k-1\\
2k & 2k
\end{array}\right]   \nonumber
\ee

The homological diagram for the $(2,2p+1)$ torus knot consists of a single zig-zag, which is a building block of homologies for more complicated knots (as stated in Conjecture \ref{conj3}), and the above matrix represents the corresponding quiver. It is also interesting that, while increasing $p$, all previously determined entries of the matrix (\ref{C-torus22p1}) remain unchanged, so that it makes sense to consider the limit $p\to\infty$ of an infinite quiver.

Furthermore, from (\ref{eq:change of vars general}) we find parameters that determine (\ref{PxC}), which are indeed consistent with our conjectures. In particular ${\alpha_{1}}=d_{3}+d_{5}+\ldots+d_{2p+1}$ gives rise to the minus sign $(-1)^{\alpha_1}$ in (\ref{PxC}), which is consistent with the sign $(-1)^{\sum_i t_i d_i}$ determined by homological degrees $t_i$, encoded in the diagonal of (\ref{C-torus22p1})
\be
(t_i) = (0,2,3,4,5,\ldots,2p,2p+1).
\ee
In addition, the parameters $a_i$ and $l_i$ (and so $q_i$) in (\ref{PxC}) are determined by
\be
\sum_i a_{i} d_i =  2pr+2\alpha_{1}=2p(d_{1}+d_{2}+d_{4}+\ldots+d_{2p}) +2(p+1)(d_{3}+d_{5}+\ldots+d_{2p+1}), 
\ee
\begin{align}
\sum_{i}l_{i} d_{i} &= -2pr+2(k_{1}+k_{2}+\ldots+k_{p})-3\alpha_{1} = \nonumber \\
& = -2pd_{1}+2(1-p)d_{2}+\big(2(1-p)-3\big)d_{3} +\nonumber \\
 &\quad +2(2-p)d_{4}+\big(2(2-p)-3\big)d_{5} +\nonumber \\
 &\qquad \quad \vdots\\
 &\quad +2(p-1-p)d_{2(p-1)}+\big(2(p-1-p)-3\big)d_{2(p-1)+1} + \nonumber \\
 &\quad +2(p-p)d_{2p}+\big(2(p-p)-3\big)d_{2p+1}.\nonumber 
\end{align}

As a confirmation, for trefoil and cinquefoil knots, restricting (\ref{C-torus22p1}) to $p=1$ and $p=2$, we reproduce respectively (\ref{trefoil-quiver}) and (\ref{51-quiver})
\begin{equation}
C^{T_{2,3}}=\left[\begin{array}{cc}
F_{0} & F_{1}\\
F_{1}^{T} & D_{1}
\end{array}\right]=\left[\begin{array}{ccc}
0 & 1 & 1\\
1 & 2 & 2\\
1 & 2 & 3
\end{array}\right] \qquad \qquad
C^{T_{2,5}}=\left[\begin{array}{ccc}
F_{0} & F_{1} & F_{2}\\
F_{1}^{T} & D_{1} & U_{2}\\
F_{2}^{T} & U_{2}^{T} & D_{2}
\end{array}\right]=\left[\begin{array}{ccccc}
0 & 1 & 1 & 3 & 3\\
1 & 2 & 2 & 3 & 3\\
1 & 2 & 3 & 4 & 4\\
3 & 3 & 4 & 4 & 4\\
3 & 3 & 4 & 4 & 5
\end{array}\right]
\end{equation}


\subsection{$(2,2p)$ torus links}

In general, the analysis of HOMFLY-PT homology of links is more involved. However if all components of a link are colored by the same representation, they have properties analogous to knots. In particular colored HOMFLY-PT polynomials for $(2,2p)$ torus links, with all components colored by the same symmetric representation $S^r$, take the form \cite{Gukov:2015gmm}
\begin{align}
P_{[r]}^{T_{2,2p}}(a,q) &=  a^{2pr}q^{-2pr}\sum_{0\leq s_{1}\leq\ldots\leq s_{p}\leq s_{p+1}=r}(a^{2}q^{-2};q^{2})_{s_{p}}(q^{2};q^{2})_{r-s_{1}}\times\nonumber \\
 & \quad \times \prod_{i=1}^{p}q^{4s_{i}}(-q)^{-2s_{i}}q^{2rs_{i}-s_{i}s_{i+1}}{ s_{i+1}\brack  s_{i}} . \label{eq:Torus link formula from GNSSS}
\end{align}
This expression corresponds to the so-called "finite-dimensional" version, which is a suitably normalized reduced colored HOMFLY-PT polynomial, that is actually a polynomial. It can be also rewritten as 
\begin{align}
P_{[r]}^{T_{2,2p}}(a,q) &= a^{2pr}q^{-2pr}\sum_{0\leq k_{p}\leq\ldots\leq k_{1}\leq k_{0}=r} {r\brack k_{1}}
{k_{1}\brack  k_{2}} \cdots {k_{p-1}\brack  k_{p}} \times\nonumber \\
 & \quad \times q^{2\sum_{i=1}^{p}((2r+1)k_{i}-k_{i-1}k_{i})}(a^{2}q^{-2};q^{2})_{k_{1}}(q^{2};q^{2})_{r-k_{p}}.\label{eq:P_r for general torus link}
\end{align}
Following analogous manipulations as in section \ref{ssec-torus22p1} we find that this expression can be further rewritten in the form (\ref{PxC}), with the corresponding quiver encoded in the matrix
\begin{equation}
C^{T_{2,2p}}=\left[\begin{array}{ccccccc}
F_{0} & F_{1} & F_{2} & F_{3} & \cdots & F_{p-1} & F_{p}^{e}\\
F_{1}^{T} & D_{1} & U_{2} & U_{3} & \cdots & U_{p-1} & U_{p}^{e}\\
F_{2}^{T} & U_{2}^{T} & D_{2} & U_{3} & \cdots & U_{p-1} & U_{p}^{e}\\
F_{3}^{T} & U_{3}^{T} & U_{3}^{T} & D_{3} & \cdots & U_{p-1} & U_{p}^{e}\\
\vdots & \vdots & \vdots & \vdots & \ddots & \vdots & \vdots\\
F_{p-1}^{T} & U_{p-1}^{T} & U_{p-1}^{T} & U_{p-1}^{T} & \cdots & D_{p-1} & U_{p}^{e}\\
F_{p}^{eT} & U_{p}^{eT} & U_{p}^{eT} & U_{p}^{eT} & \cdots & U_{p}^{eT} & D_{p}^{e}
\end{array}\right]    \label{C-quiver22p}
\end{equation}
Apart from the last column and row, the block entries take the form
\be
F_{0}=\left[\begin{array}{cc}
0 & 0\\
0 & 1
\end{array}\right], \qquad \qquad
F_{k}=\left[\begin{array}{cccc}
2k-1 & 2k-1 & 2k-1 & 2k-1\\
2k & 2k & 2k & 2k
\end{array}\right],
\ee
and
\be
D_{k}=\left[\begin{array}{cccc}
2k+1 & 2k+1 & 2k & 2k+1\\
2k+1 & 2k+2 & 2k & 2k+1\\
2k & 2k & 2k & 2k    \\
2k+1 & 2k+1 & 2k & 2k+1
\end{array}\right],\qquad 
U_{k}=\left[\begin{array}{cccc}
2k & 2k & 2k & 2k   \\
2k+1 & 2k+1 & 2k+1 & 2k+1\\
2k-1 & 2k-1 & 2k-1 & 2k-1\\
2k & 2k & 2k & 2k
\end{array}\right]
\ee
In addition, the terms in the last column and row take the form
\be
F_{p}^{e}=\left[\begin{array}{cc}
2p-1 & 2p-1\\
2p-1 & 2p-1
\end{array}\right], \qquad
D_{p}^{e}=\left[\begin{array}{cc}
2p+1 & 2p\\
2p & 2p
\end{array}\right], \qquad
U_{p}^{e}=\left[\begin{array}{cc}
2p & 2p\\
2p & 2p\\
2p-1 & 2p-1\\
2p-1 & 2p-1
\end{array}\right]
\ee
The matrix (\ref{C-quiver22p}), being assigned to a link with two components, in fact represents a combination of two (appropriately shifted) zig-zags (\ref{C-torus22p1}).

Furthermore, the linear terms that determine (\ref{PxC}) take the form
\begin{align}
(-1)^{\sum_{i} t_{i} d_{i}}&= (-1)^{ (d_{3}+d_{4})+(d_{7}+d_{8})+\ldots+(d_{4p-1}+d_{4p}) +2 (d_{5}+d_{9}+\ldots+d_{4p+1} )}, \nonumber \\
\sum_{i}a_{i} d_{i}& =  2p\big(d_{1}+(d_{2}+d_{3})+(d_{6}+d_{7})+\ldots+(d_{4(p-1)-2}+d_{4(p-1)-1})+d_{4p+1}\big)+ \nonumber \\
 &\quad +2(p+1)\big( (d_{4}+d_{5})+(d_{8}+d_{9})+\ldots+(d_{4(p-1)}+d_{4(p-1)+1})\big), \\
\sum_{i}l_{i} d_{i} &=  (-2p)d_{1}+(-2p)d_{2}+(-1-2p)d_{3}+(-1-2p)d_{4}+(-2-2p)d_{5} + \nonumber \\
 &\quad +(2-2p)d_{6}+(1-2p)d_{7}+(1-2p)d_{8}+(-2p)d_{9} + \nonumber \\
 &\quad +(4-2p)d_{10}+(3-2p)d_{11}+(3-2p)d_{12}+(2-2p)d_{13} + \nonumber \\
 &\qquad \quad \vdots\\
 &\quad -4d_{4(p-1)-2}-5d_{4(p-1)-1}-5d_{4(p-1)}-6d_{4(p-1)+1} + \nonumber \\
 &\quad -2d_{4p-2}-3d_{4p-1}-3d_{4p}.\nonumber 
\end{align}

Specializing (\ref{C-quiver22p}) we find, for example, that quivers for the Hopf link ($p=1$) and the $T_{2,4}$ link ($p=2$) are represented by matrices
\begin{equation}
C^{T_{2,2}}=\left[\begin{array}{cc}
F_{0} & F_{1}^{e}\\
F_{1}^{eT} & D_{1}^{e}
\end{array}\right]=\left[\begin{array}{cccc}
0 & 0 & 1 & 1\\
0 & 1 & 1 & 1\\
1 & 1 & 3 & 2\\
1 & 1 & 2 & 2
\end{array}\right] \qquad
C^{T_{2,4}}=\left[\begin{array}{ccc}
F_{0} & F_{1} & F_{2}^{e}\\
F_{1}^{T} & D_{1} & U_{2}^{e}\\
F_{2}^{eT} & U_{2}^{eT} & D_{2}^{e}
\end{array}\right]=\left[\begin{array}{cccccccc}
0 & 0 & 1 & 1 & 1 & 1 & 3 & 3\\
0 & 1 & 2 & 2 & 2 & 2 & 3 & 3\\
1 & 2 & 3 & 3 & 2 & 3 & 4 & 4\\
1 & 2 & 3 & 4 & 2 & 3 & 4 & 4\\
1 & 2 & 2 & 2 & 2 & 2 & 3 & 3\\
1 & 2 & 3 & 3 & 2 & 3 & 3 & 3\\
3 & 3 & 4 & 4 & 3 & 3 & 5 & 4\\
3 & 3 & 4 & 4 & 3 & 3 & 4 & 4
\end{array}\right]
\end{equation}
The matrix $C^{T_{2,2}}$ for the Hopf-link represents two zig-zags, respectively of length 3 (which is identical to a matrix for the trefoil knot (\ref{trefoil-quiver})) and of length 1 (representing a single homology generator of $t$-degree 1). The matrix $C^{T_{2,4}}$ for the $T_{2,4}$ link consists of one zig-zag of length 5 (identical to a matrix for the $5_1$ knot (\ref{51-quiver})), and another zig-zag of length 3 (identical to a matrix for the trefoil knot, but with all elements shifted by 1, with $(1,3,4)$ on the diagonal).


\subsection{$(3,p)$ torus knots}

We discuss now torus knots from the $(3,p)$ family, which enables us to present other interesting aspects of the duality with quivers. Properties of these knots are much more involved than for $(2,2p+1)$ torus knots, in particular their homology is thick. General formulas for colored superpolynomials for arbitrary $(3,p)$ torus knot are unknown. Although there are explicit expressions for the colored HOMFLY-PT polynomials of arbitrary colors for torus knots, via Rosso-Jones formula \cite{RossoJones:1993}, the formulas involve different plethysm coefficients, that are changing with the colors. In such a way they are not suitable for obtaining explicit expressions for arbitrary symmetric color and consequently for obtaining the explicit generating function of colored HOMFLY-PT polynomials for general torus knots that we need for our main Conjecture \ref{PxCConj}. 
However colored superpolynomials for the special cases of $(3,4)$ and $(3,5)$ torus knots (equivalently, respectively, $8_{19}$ and $10_{124}$ knots) were determined in \cite{Gukov:2015gmm}. In what follows we show, first, that these formulas can be rewritten in the general quiver form, and we identify corresponding quivers. Second, we show that such quivers are determined not uniquely, but only up to a permutation of some of their entries, which indicates some symmetry of the corresponding quiver moduli spaces. Furthermore, by simply assuming that there should exist a corresponding quiver, we find explicit formulas for colored HOMFLY-PT polynomials of $(3,7)$ torus knot, which have not been known before, and which nicely illustrate the power of our formalism. 

Here we focus only on $(3,p)$ knots, rather than links, so $p$ cannot be a multiple of 3. The case of $p=1$ is the framed unknot, and $p=2$ represents the trefoil, already analyzed in section \ref{ssec-3151}. Therefore the first nontrivial examples involve $p=3,4$ and 7. Moreover, as computations become more involved and technical, in this section we only consider extremal (bottom row) invariants (\ref{quivgen2}); with some patience, and taking advantage of structural properties presented in Conjecture \ref{conj3}, these results can be generalized to the full $a$-dependence.

Let us consider the $(3,4)$ torus knot first. Its quadruply-graded Poincar{\'e} polynomial determined in \cite{Gukov:2015gmm} reads
\begin{align}
P_r(a,Q,t_r,t_c) &=  a^{6 r} Q^{6 r} t_c^{6 r^2} t_r^{6 r} \sum _{\alpha=0}^r \sum _{\beta=\alpha}^r \sum _{\gamma=\beta}^r \sum _{j=\gamma}^r a^{2 (j-\gamma)} {\beta \brack \alpha}_{t_c^{-2}} {\gamma \brack \beta}_{t_c^{-2}} {j \brack \gamma}_{t_c^{-2}} {r \brack j}_{t_c^{-2}} \times \nonumber \\
& \times Q^{-4 \alpha-4 \beta+4 \gamma-8 j} t_c^{-2 (\alpha^2+\beta^2+\gamma^2 )-2 (j-\gamma) (\alpha+\beta+\gamma)-(j-\gamma)^2} 
t_r^{-2 (\alpha+\beta+\gamma)-(j-\gamma)} \times  \\ 
& \times \Big(-\frac{Q^2}{a^2 t_r t_c}; {t_c}^{-2}\Big)_{j-\gamma} \big(-a^2 Q^2 t_r^3 t_c^{2 r+1};t_c^2\big)_j,  \nonumber
\end{align}
where now we denote ${\beta \brack \alpha}_{t_c^{-2}}=\frac{(t_c^{-2};t_c^{-2})_{\beta}}{ (t_c^{-2};t_c^{-2})_{\alpha} (t_c^{-2};t_c^{-2})_{\beta-\alpha} }$. Upon the identification of variables (\ref{super-homfly}), and extracting the terms at the lowest powers of $a$, we find that that extremal (bottom row) colored HOMFLY-PT polynomials for the $(3,4)$ torus knot take form
\begin{align}
P_r^{bottom}(q) = q^{6 r^2} \sum _{\alpha=0}^r \sum_{\beta=\alpha}^r \sum _{\gamma=\beta}^r \sum _{j=\gamma}^r \frac{(q^2;q^2)_r \,q^{-2\alpha (\beta-\gamma+j+1)-2\beta (j+1)+2\gamma- 2j (r+2)}}{(q^2;q^2)_{\alpha} (q^2;q^2)_{\beta-\alpha} (q^2;q^2)_{\gamma-\beta}  \label{P34-bottom} (q^2;q^2)_{j-\gamma} (q^2;q^2)_{r-j}},
\end{align}
while its uncolored HOMFLY-PT homology has 5 generators in the bottom row, whose $q$-degrees and $t$-degrees are
\be
\begin{split}
(q_1,q_2,q_3,q_4,q_5)&=(-6,-2,0,2,6),   \\
(t_1,t_2,t_3,t_4,t_5)&=(0,2,4,6,8).   \label{T34-qt}
\end{split}
\ee
Manipulating the expression (\ref{P34-bottom})  we find that the corresponding quiver is represented by the following matrix
\begin{equation}
C^{T_{3,4}}=\left[\begin{array}{ccccc}
0&1&2&3&5\\
1&2&3&3&5\\
2&3&4&4&5\\
3&3&4&4&5\\
5&5&5&5&6
\end{array}\right]
\end{equation}
This quiver, together with $q$-degrees and $t$-degrees of 5 bottom row generators in (\ref{T34-qt}), encode all extremal (bottom row) colored HOMFLY-PT polynomials for the $(3,4)$ torus knot, which can be reconstructed from (\ref{quivgen2}). Moreover, simply the fact that we are able to identify this quiver proves the LMOV conjecture for all symmetric representations for this knot. Furthermore, the matrix $C^{T_{3,4}}$ captures the structure of the bottom row generators of HOMFLY-PT homology for the $(3,4)$ torus knot, which consists of one zig-zag (the same as for the $(2,7)$ torus knot) and one diamond. The part of the matrix with $(0,2,4,6)$ on the diagonal represents the bottom row of the zig-zag, and an additional 4 on the diagonal is the bottom row of the diamond (\ref{diamond}).

The next knot in this series is the $(3,5)$ torus knot. Its HOMFLY-PT homology and colored superpolynomials were also considered in \cite{Gukov:2015gmm}. For brevity, we just recall that colored superpolynomials for this knot take form
\begin{align}
P_r&(a,q,t) = \sum_{j=0}^r \sum_{k_1 = 0}^j \sum_{k_2 = 0}^{k_1} \sum_{k_3 = 0}^{k_2} \sum_{k_4 = 0}^{k_3} \sum_{i=0}^{r-j} a^{8r} \bigg(\frac{t}{q}\bigg)^{2(i+2j - k_1 - k_2 -k_3 -k_4 + 2r)}   \times   \\
& \times q^{-2(k_1 k_2 + k_2 k_3 + k_3 k_4 + 2 (k_1+k_2+k_3+k_4)+r+(k_1+k_2+k_3+k_4)r-2r^2-i(2+k_1+r) -j(5+k_2+k_3+k_4+2r)}  \times \nonumber \\
& \times {j \brack k_1} {k_1 \brack k_2} {k_2\brack k_3} {k_3\brack k_4} {r-j \brack i} 
 (-a^2 t q^{-2};q^2)_{r-j} (-a^2 q^{-2-2j+2r};q^2)_{k_4 - j} (-a^2 q^{2r} t^3;q^2)_{r-j}.  \nonumber
\end{align}
Setting $t=-1$ and extracting coefficients of minimal powers of $a$ reveals the form of extremal (bottom row) colored HOMFLY-PT polynomials
\begin{align} 
P&^{bottom}_r(q) =  \sum_{j=0}^r \sum_{k_1 = 0}^j \sum_{k_2 = 0}^{k_1} \sum_{k_3 = 0}^{k_2} \sum_{k_4 = 0}^{k_3} \sum_{i=0}^{r-j}\  (q^2;q^2)_r   \times \\
&\times \frac{q^{-2 (k_2+k_3+k_4)-2 (k_1+k_1 k_2+k_2 k_3+k_3 k_4)-6 r-2 (k_1+k_2+k_3+k_4) r+4 r^2+2 i (1+k_1+r)+2 j (3+k_2+k_3+k_4+2 r)}}{(q^2;q^2)_i (q^2;q^2)_{j-k_1} (q^2;q^2)_{k_1-k_2} (q^2;q^2)_{k_2-k_3} (q^2;q^2)_{k_3-k_4} (q^2;q^2)_{k_4} (q^2;q^2)_{r-j-i}}.  \nonumber
 \end{align}
Furthermore, the HOMFLY-PT homology for $(3,5)$ torus knot has 7 generators in the bottom row, with the following $q$-degrees and $t$-degrees
\be
\begin{split}
(q_1,q_2,\ldots,q_7)&=(-8,-4,-2,0,2,4,8),   \\
(t_1,t_2,\ldots,t_7)&=(0,2,4,4,6,6,8).    \label{P35-qt}
\end{split}
\ee
Manipulating the above expressions we find, surprisingly, that there are two quivers which reproduce the same extremal (bottom row) colored HOMFLY-PT polynomials via (\ref{quivgen2}). These two quivers are very similar, and their matrices differ only by a permutation of a few entries. They take the following form; the permuted entries are underlined in the second matrix
\begin{equation}
C^{T_{3,5}}=\left[\begin{array}{ccccccc}
0&1&2&3&4&5&7\\
1&2&3&3&5&5&7\\
2&3&4&4&5&5&7\\
3&3&4&4&6&5&7\\
4&5&5&6&6&6&7\\
5&5&5&5&6&6&7\\
7&7&7&7&7&7&8
\end{array}\right] \qquad \qquad
\widetilde{C}^{T_{3,5}}=\left[\begin{array}{ccccccc}
0&1&2&3&4&5&7\\
1&2&3&3&5&5&7\\
2&3&4&4&5&5&\underline{6}\\
3&3&4&4&6&5&7\\
4&5&5&6&6&\underline{7}&7\\
5&5&5&5&\underline{7}&6&7\\
7&7&\underline{6}&7&7&7&8
\end{array}\right]
\end{equation}
Again let us stress, that the quiver represented by either of the above matrcies, together with $q$-degrees and $t$-degrees in (\ref{P35-qt}), encode all extremal colored HOMFLY-PT polynomials for $(3,5)$ torus knot, which can be reconstructed from (\ref{quivgen2}). The fact that we are able to identify these quivers also proves the LMOV conjecture for all symmetric representations for this knot.

The next knot we consider is the $(3,7)$ torus knot. Its colored superpolynomials, or even explicit, closed-form expressions of HOMFLY-PT polynomials colored by arbitrary symmetric representations, have not been known before. However, based on the structure (\ref{quivgen2}), and by comparing the results with the Rosso-Jones formula for the first several symmetric representations, we are able to reconstruct the corresponding quiver, which then encodes HOMFLY-PT polynomials colored by arbitrary symmetric representations, and corresponding integral LMOV invariants. The homology of the bottom row of the $(3,7)$ torus knot has 12 generators with the following $q$-degrees and $t$-degrees
\be
\begin{split}
(q_1,q_2,\ldots,q_{12})&=(-12,-8,-6,-4,-2,0,0,2,4,6,8,12),   \\
(t_1,t_2,\ldots,t_{12})&=(0,2,4,4,6,6,8,8,8,10,10,12), 
\end{split}
\ee
and we find that the corresponding quiver is encoded in a matrix
\begin{equation}
C^{T_{3,7}}=\left[\begin{array}{cccccccccccc}
0&1&2&3&4&5&5&6&7&8&9&11\\
1&2&3&3&5&5&6&7&7&9&9&11\\
2&3&4&4&5&6&6&7&8&9&10&11\\
3&3&4&4&5&5&7&7&7&9&9&11\\
4&5&5&5&6&6&7&7&8&9&10&11\\
5&5&6&5&6&6&8&7&7&9&9&11\\
5&6&6&7&7&8&8&8&9&9&10&11\\
6&7&7&7&7&7&8&8&8&9&10&11\\
7&7&8&7&8&7&9&8&8&9&9&11\\
8&9&9&9&9&9&9&9&9&10&10&11\\
9&9&10&9&10&9&10&10&9&10&10&11\\
11&11&11&11&11&11&11&11&11&11&11&12
\end{array}\right].
\end{equation}
In this case we also find other quivers, whose matrices differ from the above one by permutation of several entries, and yet encode the same generating series (\ref{quivgen2}).


\subsection{Twist knots $4_1, 6_1, 8_1, \ldots$}

Another infinite family of knots that we consider are twist knots, which are labelled by an integer $p$. Negative values of this parameter, i.e. $p=-1,-2,-3,\ldots,$ correspond to $4_1, 6_1,8_1,\ldots$ knots; these are simply $(2|p|+2)_1$ knots, which are also denoted $TK_{2|p|+2}$. The colored HOMFLY-PT polynomials for these knots are determined in \cite{FGSS,Nawata} and they take the form
\begin{align}
P_{r}^{TK_{2|p|+2}}(a,q)= & \sum_{0\leq k_{|p|}\leq\ldots\leq k_{2}\leq k_{1}\leq r}{r \brack k_1}{k_1 \brack k_2}\cdots{k_{|p|-1}\brack k_{|p|}}\times\nonumber \\
 & \qquad \times a^{2\sum_{i=1}^{|p|}k_{i}}q^{2\sum_{i=1}^{|p|}\left(k_{i}^{2}-k_{i}\right)}(a^{-2}q^{2};q^{-2})_{k_{1}}(a^{-2}q^{-2r};q^{-2})_{k_{1}}.       \label{eq:P_r for general twist knot with p<0}
\end{align}
Following manipulations similar to the previous examples, we find that the quiver corresponding to a given $p<0$, i.e. to a given $TK_{2|p|+2}$ knot, is encoded in the matrix
\begin{equation}
C^{TK_{2|p|+2}}=\left[\begin{array}{ccccccc}
F_{0} & F & F & F & \cdots & F & F\\
F^{T} & D_{1} & R_{1} & R_{1} & \cdots & R_{1} & R_{1}\\
F^{T} & R_{1}^{T} & D_{2} & R_{2} & \cdots & R_{2} & R_{2}\\
F^{T} & R_{1}^{T} & R_{2}^{T} & D_{3} & \cdots & R_{3} & R_{3}\\
\vdots & \vdots & \vdots & \vdots & \ddots & \vdots & \vdots\\
F^{T} & R_{1}^{T} & R_{2}^{T} & R_{3}^{T} & \cdots & D_{|p|-1} & R_{|p|-1}\\
F^{T} & R_{1}^{T} & R_{2}^{T} & R_{3}^{T} & \cdots & R_{|p|-1}^{T} & D_{|p|}
\end{array}\right]
\end{equation}
where
\be
F_{0}=\left[0\right]\qquad\qquad F=\left[\begin{array}{cccc}
0 & -1 & 0 & -1\end{array}\right]
\ee
and
\be
D_{k}=\left[\begin{array}{cccc}
2k & 2k-2 & 2k-1 & 2k-3\\
2k-2 & 2k-3 & 2k-2 & 2k-4\\
2k-1 & 2k-2 & 2k-1 & 2k-3\\
2k-3 & 2k-4 & 2k-3 & 2k-4
\end{array}\right] \qquad
R_{k}=\left[\begin{array}{cccc}
2k & 2k-2 & 2k-1 & 2k-3\\
2k-1 & 2k-3 & 2k-2 & 2k-4\\
2k & 2k-1 & 2k-1 & 2k-3\\
2k-2 & 2k-3 & 2k-2 & 2k-4
\end{array}\right]
\ee
The element $F_0$ represents a zig-zag of length 1, corresponding to a single homology generator, while the diagonal blocks  $D_k$ represent (up to a permutation of homology generators, and an overall shift) diamonds (\ref{diamond}).
Note that in this case it also makes sense to consider the $-p\to\infty$ limit, and the corresponding infinite quiver.

The other parameters that determine the form of the generating series (\ref{PxC}) for the $TK_{2|p|+2}$ knot take the form
\begin{align}
(-1)^{\sum_{i}p_{i} d_{i}} &= (-1)^{ (d_{3}+d_{4})+(d_{7}+d_{8})+\ldots+(d_{4|p|-1}+d_{4|p|}) +2(d_{5}+d_{9}+\ldots+d_{4|p|+1})  } , \nonumber \\
\sum_{i}a_{i} d_{i}& =  2d_{2}+0d_{3}+0d_{4}-2d_{5}+\nonumber \\
 & \qquad +4d_{6}+2d_{7}+2d_{8}+0d_{9}+\\
 & \qquad \quad \vdots\nonumber \\
 & \qquad +2|p|d_{4|p|-2}+\left(2|p|-2\right)d_{4|p|-1}+\left(2|p|-2\right)d_{4|p|}+\left(2|p|-4\right)d_{4|p|+1}, \nonumber \\
\sum_{i}l_{i} d_{i} &=  -2d_{2}-d_{3}+d_{4}+2d_{5} + \nonumber \\
 & \qquad -4d_{6}-3d_{7}-1d_{8}+0d_{9}+\nonumber \\
 & \qquad \quad \vdots\\
 & \qquad -2|p|d_{4|p|-2}+\left(1-2|p|\right)d_{4|p|-1}+\left(3-2|p|\right)d_{4|p|}+\left(4-2|p|\right)d_{4|p|+1}.\nonumber 
\end{align}

For example, the quiver for the $p=-1$ case, i.e. the figure-eight knot $4_1$, whose homology diagram is shown in figure \ref{diagrams}, is represented by the matrix
\begin{equation}
C^{TK_{4}}=\left[\begin{array}{cc}
F_{0} & F\\
F^{T} & D_{1}
\end{array}\right]=\left[\begin{array}{ccccc}
0 & 0 & -1 & 0 & -1\\
0 & 2 & 0 & 1 & -1\\
-1 & 0 & -1 & 0 & -2\\
0 & 1 & 0 & 1 & -1\\
-1 & -1 & -2 & -1 & -2
\end{array}\right]
\end{equation}
This matrix is consistent with Conjecture \ref{conj3}: the top left entry $0$ represents the zig-zag of length 1, and the remaining diagonal block of size $4\times 4$ represents a diamond and agrees (up to a permutation of homology generators, and corresponding rows and columns) with (\ref{diamond}) for $k=-2$.

For $p=-2$, i.e. the $6_{1}$ knot, the quiver is represented by the following matrix
\begin{equation}
C^{TK_{6}}=\left[\begin{array}{ccc}
F_{0} & F & F\\
F^{T} & D_{1} & R_{1}\\
F^{T} & R_{1}^{T} & D_{2}
\end{array}\right]=\left[\begin{array}{ccccccccc}
0 & 0 & -1 & 0 & -1 & 0 & -1 & 0 & -1\\
0 & 2 & 0 & 1 & -1 & 2 & 0 & 1 & -1\\
-1 & 0 & -1 & 0 & -2 & 1 & -1 & 0 & -2\\
0 & 1 & 0 & 1 & -1 & 2 & 1 & 1 & -1\\
-1 & -1 & -2 & -1 & -2 & 0 & -1 & 0 & -2\\
0 & 2 & 1 & 2 & 0 & 4 & 2 & 3 & 1\\
-1 & 0 & -1 & 1 & -1 & 2 & 1 & 2 & 0\\
0 & 1 & 0 & 1 & 0 & 3 & 2 & 3 & 1\\
-1 & -1 & -2 & -1 & -2 & 1 & 0 & 1 & 0
\end{array}\right]         \label{C-61}
\end{equation}
Note that in these examples some entries of matrices $C$ are negative. In order to have a proper quiver representation theory interpretation, we can change the framing (\ref{C-frame}) to shift all values of these matrices by a constant and make them nonnegative.


\subsection{Twist knots $3_1, 5_2, 7_2, 9_2, \ldots$}

Another class of twist knots is characterized by $p>0$, which are respectively $3_1, 5_2,7_2,9_2,\ldots$ knots, which have $2p+1$ crossings, and are also denoted $TK_{2p+1}$. Their superpolynomials take the form \cite{Nawata}
\begin{align}
P_{S^{n-1}}^{TK_{2p+1}}(a,q,t)= & \sum_{0\leq s_{1}\leq\ldots\leq s_{p}<\infty}(-t)^{-n+1}q^{2s_{p}}\frac{(-a^{2}tq^{-2};q^{2})_{s_{p}}}{(q^{2};q^{2})_{s_{p}}}(q^{2-2n};q^{2})_{s_{p}}(-a^{2}t^{3}q^{2n-2};q^{2})_{s_{p}}\times\nonumber \\
 & \qquad\times\prod_{i=1}^{p-1}q^{4s_{i}}(a^{2}t^{2})^{s_{i}}q^{2s_{i}(s_{i}-1)}{s_{i+1}\brack s_{i}}.  \label{eq:Twist knot with p>0 formula from NRZ}
\end{align}

Here we can illustrate another subtlety, which is the fact that sometimes -- in particular for $TK_{2p+1}$ knots -- more general quivers can be assigned to a given knot, which are however not consistent with our conjectures. For example, setting $p=1$ and $t=-1$ in (\ref{eq:Twist knot with p>0 formula from NRZ}), we find the following representation of the colored HOMFLY-PT polynomials for the trefoil knot
\begin{equation}
P_{S^{r}}^{TK_{3}}(a,q)=\sum_{0\leq s_{1}\leq r} {r\brack s_{1}} (-1)^{s_{1}}q^{-2s_{1}r+s_{1}^{2}+s_{1}}(a^{2}q^{-2};q^{2})_{s_{1}}(a^{2}q^{2r};q^{2})_{s_{1}}. \label{eq:Trefoil knot treated as twist formula}
\end{equation}
This expression is equal to (\ref{Pr-31-homfly}), however its naive rewriting in the form (\ref{PxC}) leads to the quiver represented by the following matrix
\begin{equation}
\left[\begin{array}{ccccc}
0 & -1 & 0 & -1 & 0\\
-1 & -1 & 0 & -1 & 0\\
0 & 0 & 2 & 1 & 2\\
-1 & -1 & 1 & 0 & 1\\
0 & 0 & 2 & 1 & 3
\end{array}\right]
\end{equation}
The last three rows and columns of this matrix contain the trefoil quiver matrix (\ref{trefoil-quiver}) that we found earlier, however now we find two additional rows and columns. In fact the same issue arises for all twist knots $TK_{2p+1}$ in this series. Moreover, the structure of terms with linear powers of $d_i$ in the generating series that determines such enlarged quivers,  is also not quite consistent with the structure of the parameters in (\ref{PxC}) and their relation to homological degrees. Nonetheless, we can get rid of these additional rows and columns, at the same time fixing terms with linear powers of $d_i$, by taking advantage of the following lemma.

\begin{lemma}

Consider a generating function (not necessarily related to a knot) of the form (\ref{PxC}), determined by a  quiver $C$ of size $n\times n$. Up to appropriate adjustment of terms with with linear powers of $d_i$, the same generating function is assigned to the modified quiver
\begin{equation}
C^{+}=\left[\begin{array}{ccccccc}
1+\alpha_{0} & \alpha_{0} & \alpha_{1} & \alpha_{2} & \cdots & \alpha_{n-1} & \alpha_{n}\\
\alpha_{0} & \alpha_{0} & \alpha_{1} & \alpha_{2} & \cdots & \alpha_{n-1} & \alpha_{n}\\
\alpha_{1} & \alpha_{1}\\
\alpha_{2} & \alpha_{2}\\
\vdots & \vdots &  &  & C\\
\alpha_{n-1} & \alpha_{n-1}\\
\alpha_{n} & \alpha_{n}
\end{array}\right]\label{eq:C+}
\end{equation}
for every $\alpha_{0},\alpha_{1},\ldots,\alpha_{n}\in\mathbb{Z}$.

\end{lemma}

\textbf{Proof:}\\

Note that, for $m\geq 1$, using (\ref{qpoch-sum}), we have
\begin{equation}
0=(1;q^{-2})_{m}=\sum\limits _{a+b=m}(-1)^{b}q^{-b^{2}+b-2ab}\frac{(q^2;q^2)_m}{(q^2;q^2)_a (q^2;q^2)_b}.
\end{equation}
It follows that, for arbitrary $\alpha_{0},\alpha_{1},\ldots,\alpha_{n},d_{1},\ldots,d_{n}\in\mathbb{Z}$,
\begin{align}
1 & = \sum\limits _{m\geq0}q^{(\alpha_{0}+1)m^{2}+2(\alpha_{1}d_{1}+\alpha_{2}d_{2}+\ldots+\alpha_{n}d_{n})m} x^m\frac{(1;q^{-2})_{m}}{(q^{2};q^{2})_{m}} = \nonumber \\
&= \sum\limits _{a,b\geq0}(-1)^{b}q^{-b^{2}+b-2ab+(\alpha_{0}+1)(a+b)^{2}+2(\alpha_{1}d_{1}+\alpha_{2}d_{2}+\ldots+\alpha_{n}d_{n})(a+b)}\frac{x^{a+b}}{(q^{2};q^{2})_{a}(q^{2};q^{2})_{b}}.  \label{1ab} 
\end{align}
Therefore, if $P_{C}$ is a generating function determined by a quiver $C$, then
\begin{equation}
P_{C^{+}}=\sum\limits _{a,b\geq0}q^{-b^{2}+b-2ab+(\alpha_{0}+1)(a+b)^{2}+2(\alpha_{1}d_{1}+\alpha_{2}d_{2}+\ldots+\alpha_{n}d_{n})(a+b)}\frac{(-1)^{b} x^{a+b}}{(q^{2};q^{2})_{a}(q^{2};q^{2})_{b}}P_{C}\label{eq:auxiliarity of the first two variables}
\end{equation}
is a generating function determined by a quiver $C^{+}$ in (\ref{eq:C+}), and from (\ref{1ab}) we clearly see that $P_{C^{+}}=P_{C}$, which completes the proof.   \kraj

Having in mind the above subtlety, in order to find a quiver representation consistent with our conjectures, we rewrite the colored HOMFLY-PT polynomials (\ref{eq:Twist knot with p>0 formula from NRZ}) in the form
\begin{align}
P_{r}^{TK_{2p+1}}(a,q)= & \sum_{0\leq k_{p}\leq\ldots\leq k_{2}\leq k_{1}\leq r} {r\brack k_{1}} {k_{1}\brack k_{2}}\cdots {k_{p-1}\brack k_{p} } \times\nonumber \\
 & \qquad \times a^{2\sum_{i=2}^{p}k_{i}}q^{-2k_{1}r+k_{1}^{2}+k_{1}+2\sum_{i=2}^{p}\left(k_{i}^{2}-k_{i}\right)}(a^{2}q^{-2};q^{2})_{k_{1}}(a^{2}q^{2r};q^{2})_{k_{1}}.\label{eq:P_r for general twist knot with p>0}
\end{align}
Following manipulations analogous to the previous sections, we now find that the generating series (\ref{PxC}) for $TK_{2p+1}$ knot is determined by a quiver whose matrix takes the form
\begin{equation}
C^{TK_{2p+1}}=\left[\begin{array}{ccccccc}
D_{1} & R_{1} & R_{1} & R_{1} & \cdots & R_{1} & R_{1}\\
R_{1}^{T} & D_{2} & R_{2} & R_{2} & \cdots & R_{2} & R_{2}\\
R_{1}^{T} & R_{2}^{T} & D_{3} & R_{3} & \cdots & R_{3} & R_{3}\\
R_{1}^{T} & R_{2}^{T} & R_{3}^{T} & D_{4} & \cdots & R_{4} & R_{4}\\
\vdots & \vdots & \vdots & \vdots & \ddots & \vdots & \vdots\\
R_{1}^{T} & R_{2}^{T} & R_{3}^{T} & R_{4}^{T} & \cdots & D_{p-1} & R_{p-1}\\
R_{1}^{T} & R_{2}^{T} & R_{3}^{T} & R_{4}^{T} & \cdots & R_{p-1}^{T} & D_{p}
\end{array}\right]     \label{C-TK2p1}
\end{equation}
where the block elements in the first row and column are
\be
D_{1}=\left[\begin{array}{ccc}
2 & 1 & 2\\
1 & 0 & 1\\
2 & 1 & 3
\end{array}\right]\qquad \qquad 
R_{1}=\left[\begin{array}{cccc}
1 & 2 & 1 & 2\\
0 & 2 & 0 & 1\\
1 & 3 & 2 & 3
\end{array}\right]    \label{D1R1-TK2p1}
\ee
and all other elements, for $k>1$, take the form
\be
D_{k}=\left[\begin{array}{cccc}
2k-3 & 2k-2 & 2k-3 & 2k-2\\
2k-2 & 2k & 2k-1 & 2k\\
2k-3 & 2k-1 & 2k-2 & 2k-1\\
2k-2 & 2k & 2k-1 & 2k+1
\end{array}\right] \qquad \quad
R_{k}=\left[\begin{array}{cccc}
2k-3 & 2k-2 & 2k-3 & 2k-2\\
2k-1 & 2k & 2k-1 & 2k  \\
2k-2 & 2k & 2k-2 & 2k-1\\
2k-1 & 2k+1 & 2k & 2k+1
\end{array}\right]
\ee
In this case $D_1$ represents a zig-zag of the same form as for the trefoil knot (\ref{trefoil-quiver}), and the $D_k$ (for $k>1$) represent (up to a permutation of homology generators and an overall constant shift) the diamonds (\ref{diamond}).


The other parameters that determine (\ref{PxC}), now with vertices of a quiver (or homology generators), and thus also the summation variables $d_i$ numbered from $3$ to $4p+1$ (after removing $d_1$ and $d_2$ using the above lemma), take the form
\begin{align}
(-1)^{\sum_{i=3}^{4p+1}t_{i} d_{i}} &= (-1)^{ (d_{3}+d_{4})+(d_{7}+d_{8})+\ldots+(d_{4p-1}+d_{4p}) +2(d_{5}+d_{9}+\ldots+d_{4p+1})}\nonumber \\
\sum_{i=3}^{4p+1}a_{i} d_{i} &=  2(d_{3}+d_{4})+4d_{5} + \nonumber \\
 & \qquad +2d_{6}+4d_{7}+4d_{8}+6d_{9} + \\
 & \qquad +4d_{10}+6d_{11}+6d_{12}+8d_{13} + \nonumber \\
 & \qquad \quad \vdots\nonumber \\
 & \qquad +2(p-1)d_{4p-2}+2pd_{4p-1}+2pd_{4p}+2(p+1)d_{4p+1}\nonumber \\
\sum_{i=2}^{4p}l_{i} d_{i} &=  -2d_{4}-3d_{5} +\nonumber \\
 &\qquad -d_{6}-2d_{7}-4d_{8}-5d_{9}+\nonumber \\
 &\qquad -3d_{10}-4d_{11}-6d_{12}-7d_{13}+\nonumber \\
 &\qquad \quad \vdots\\
 &\qquad +(1-2p)d_{4p-2}+(2-2p)d_{4p-1}+(-2p)d_{4p}+(-1-2p)d_{4p+1}.\nonumber 
\end{align}

For example, for $p=1$, which represents simply the trefoil knot, the quiver matrix (\ref{C-TK2p1}) consists only of the element $D_1$ in (\ref{D1R1-TK2p1}), and up to permutation of vertices it is equivalent to (\ref{trefoil-quiver}). For $p=2$, i.e. for the knot $5_{2}$, from (\ref{C-TK2p1}) we obtain a quiver represented by the matrix
\begin{equation}
C^{TK_{5}}=\left[\begin{array}{ccc}
D_{1} & R_{1} & R_{1}\\
R_{1}^{T} & D_{2} & R_{2}\\
R_{1}^{T} & R_{2}^{T} & D_{3}
\end{array}\right]=\left[\begin{array}{ccccccc}
2 & 1 & 2 & 1 & 2 & 1 & 2\\
1 & 0 & 1 & 0 & 2 & 0 & 1\\
2 & 1 & 3 & 1 & 3 & 2 & 3\\
1 & 0 & 1 & 1 & 2 & 1 & 2\\
2 & 2 & 3 & 2 & 4 & 3 & 4\\
1 & 0 & 2 & 1 & 3 & 2 & 3\\
2 & 1 & 3 & 2 & 4 & 3 & 5
\end{array}\right]
\end{equation}


\subsection{$6_2$ and $6_3$ knots}

Finally we discuss knots with six crossings, $6_2$ and $6_3$ (the third prime knot with six crossings is the twist knot $6_1$, whose quiver we already identified in (\ref{C-61})). Explicit expressions for colored polynomials for those knots have not been known before. Assuming that they are consistent with our conjectures, we are able to determine such expressions, as being encoded in the corresponding quivers. This again shows the power of our formalism.

Let us consider $6_2$ knot first. Its uncolored HOMFLY-PT homology has 11 generators, which have the following degrees 
\be
\begin{split}
(a_1,\ldots,a_{11})&=(0,2,2,0,2,2,2,4,4,2,4) ,   \\
(q_1,\ldots,q_{11})&=(-2,-4,-2,2,0,0,2,-2,0,4,2),    \\
(t_1,\ldots,t_{11})&=(-2,-1,0,0,1,1,2,2,3,3,4). 
\end{split}
\ee
Assuming that the generating series of colored HOMFLY-PT polynomials takes the form (\ref{PxC}), and comparing that generating series with several first such polynomials obtained using the Rosso-Jones formula, we find that the corresponding quiver is encoded in the following matrix
\be
C^{6_2}=\left[
\begin{array}{ccccccccccc}
 -2\ & -2\ & -1\ & -1\ & -1\ & -1\ & 0\ & -1\ & 1\ & 1\ & 1 \\
 -2\ & -1\ & -1\ & 0\ & 0\ & 0\ & 1\ & 0\ & 1\ & 2\ & 2 \\
 -1\ & -1\ & 0\ & 1\ & 0\ & 0\ & 1\ & 0\ & 1\ & 2\ & 2 \\
 -1\ & 0\ & 1\ & 0\ & 0\ & 0\ & 1\ & 0\ & 2\ & 1\ & 1 \\
 -1\ & 0\ & 0\ & 0\ & 1\ & 1\ & 1\ & 1\ & 2\ & 2\ & 2 \\
 -1\ & 0\ & 0\ & 0\ & 1\ & 1\ & 1\ & 1\ & 2\ & 2\ & 2 \\
 0\ & 1\ & 1\ & 1\ & 1\ & 1\ & 2\ & 1\ & 2\ & 2\ & 2 \\
 -1\ & 0\ & 0\ & 0\ & 1\ & 1\ & 1\ & 2\ & 2\ & 3\ & 3 \\
 1\ & 1\ & 1\ & 2\ & 2\ & 2\ & 2\ & 2\ & 3\ & 3\ & 3 \\
 1\ & 2\ & 2\ & 1\ & 2\ & 2\ & 2\ & 3\ & 3\ & 3\ & 3 \\
 1\ & 2\ & 2\ & 1\ & 2\ & 2\ & 2\ & 3\ & 3\ & 3\ & 4 \\
\end{array}
\right]
\ee
In this matrix one can identify diagonal blocks, one corresponding to a zig-zag of the same form as in the trefoil quiver (\ref{trefoil-quiver}), and two diamonds of the form (\ref{diamond}).

Analogously we analyze the generating series for the $6_3$ knot, whose uncolored HOMFLY-PT homology has 13 generators of the following degrees
\be
\begin{split}
(a_1,\ldots,a_{13})&=(0,2,0,0,-2,2,0,0,-2,2,0,0,-2),    \\
(q_1,\ldots,q_{13})&=(0,-2,0,-4,-2,0,2,-2,0,2,4,0,2),     \\
(t_1,\ldots,t_{13})&=(0,1,0,-2,-3,2,1,-1,-2,3,2,0,-1).
\end{split}
\ee
Similarly, comparing first few colored HOMFLY-PT polynomials from (\ref{PxC}) with the Rosso-Jones formula, we find the corresponding quiver
\be
C^{6_3}=\left[
\begin{array}{ccccccccccccc}
 0\ & 0\ & 0 & -1\ & -1\ & 0\ & 0\ & -1\ & -1\ & 0\ & 0\ & -1\ & -1 \\
 0\ & 1\ & 0 & -1\ & -2\ & 1\ & 0\ & -1\ & -2\ & 1\ & 1\ & 0\ & -1 \\
 0\ & 0\ & 0 & -1\ & -2\ & 1\ & 0\ & 0\ & -2\ & 1\ & 1\ & 0\ & 0 \\
 -1\ & -1\ & -1\ & -2\ & -3\ & 0\ & -1\ & -2\ & -3\ & -1\ & 0\ & -2\ & -2 \\
 -1\ & -2\ & -2\ & -3\ & -3\ & -1\ & -1\ & -2\ & -3\ & -1\ & -1\ & -2\ & -2 \\
 0\ & 1\ & 1\ & 0 & -1\ & 2\ & 1\ & 0\ & -1\ & 2\ & 1\ & 1\ & -1 \\
 0\ & 0\ & 0\ & -1 & -1\ & 1\ & 1\ & 0\ & -1\ & 2\ & 1\ & 1\ & 0 \\
 -1\ & -1\ & 0\ & -2 & -2\ & 0\ & 0\ & -1\ & -2\ & 0\ & 0\ & -1\ & -2 \\
 -1\ & -2\ & -2\ & -3 & -3\ & -1\ & -1\ & -2\ & -2\ & 0\ & -1\ & -1\ & -2 \\
 0\ & 1 & 1\ & -1\ & -1\ & 2\ & 2 & 0\ & 0\ & 3\ & 2\ & 1\ & 0 \\
 0\ & 1 & 1\ & 0\ & -1\ & 1\ & 1 & 0\ & -1\ & 2\ & 2\ & 1\ & 0 \\
 -1\ & 0 & 0\ & -2\ & -2\ & 1\ & 1 & -1\ & -1\ & 1\ & 1\ & 0\ & -1 \\
 -1\ & -1 & 0\ & -2\ & -2\ & -1\ & 0 & -2\ & -2\ & 0\ & 0\ & -1\ & -1 \\
\end{array}
\right]
\ee
In this matrix one can identify diagonal blocks, one corresponding to a zig-zag of length 1 (representing a homology generator with $t$-degree 0), and three diamonds of the form (\ref{diamond}).

We checked that the above results agree with those obtained using the formalism of differentials, presented in \cite{Nawata:2015wya}.




\acknowledgments{We thank Sergei Gukov, Satoshi Nawata, Mi{\l}osz Panfil, Yan Soibelman, Richard Thomas, Cumrun Vafa, and Paul Wedrich for comments and discussions. We are grateful to the American Institute of Mathematics (San Jose), the Isaac Newton Institute for Mathematical Sciences (Cambridge University), and Institut Henri Poincar{\'e} (Paris) for hospitality. The last author thanks the California Institute of Technology, the Matrix Institute (University of Melbourne, Creswick), University of California Davis, and the Isaac Newton Institute for Mathematical Sciences (Cambridge University) for hospitality and opportunity to present the results of this work during seminars and conferences within the last year. This work is supported by the ERC Starting Grant no. 335739 \emph{``Quantum fields and knot homologies''} funded by the European Research Council under the European Union’s Seventh Framework Programme, and the Foundation for Polish Science. M.S. is partially supported by the Ministry of Science of Serbia, project no. 174012.}



\newpage

\bibliographystyle{JHEP}
\bibliography{abmodel}

\end{document}